\documentclass[a4paper]{cas-sc}

\usepackage{url}
\usepackage[numbers]{natbib}
\usepackage{amssymb}
\usepackage{amsmath}
\usepackage{algpseudocode}
\usepackage{algorithm}
\usepackage{algorithmicx}
\usepackage{color,xcolor}
\usepackage{longtable}
\usepackage{float}
\usepackage{booktabs}
\usepackage{multirow} 
\usepackage{makecell}
\usepackage{graphicx}
\usepackage{hyperref}
\hypersetup{colorlinks=true, citecolor=blue, linkcolor=blue, urlcolor=blue}
\usepackage{caption}
\usepackage{enumitem}
\newtheorem{thm}{Theorem} 
\newtheorem{lem}{Lemma} 
\newdefinition{rmk}{Remark}

\def\tsc#1{\csdef{#1}{\textsc{\lowercase{#1}}\xspace}}
\tsc{WGM}
\tsc{QE}
\tsc{EP}
\tsc{PMS}
\tsc{BEC}
\tsc{DE}

\makeatletter
\let\ps@first\ps@plain
\makeatother
\ExplSyntaxOn
\RenewDocumentEnvironment{Abstract}{o}
{
  \group_begin:
  \par\medskip
  {\centering\normalfont\large\bfseries Abstract\par}
  \vspace{8pt}
  \normalfont\normalsize\noindent\ignorespaces
}
{
  \par\group_end:
}

\RenewDocumentCommand\dashrule{ O{.4pt} m m }
{
  \skip_vertical:n{#2}
  \skip_vertical:n{#3}
}
\RenewDocumentCommand\printcornotes{}{}

\RenewDocumentCommand\printemails{}{}
\ExplSyntaxOff

\usepackage{float}
\begin{document}
\hypersetup{
    colorlinks=true,
    allcolors=black
}
\let\WriteBookmarks\relax
\let\printorcid\relax 



\stmausetup{
    align=\centering
}

\stmaffsetup{
    align=\centering,
    shape=\upshape,
    size=\small,
    color=black,
    weight=\mdseries,
    before=0pt
}

\shortauthors{Ge et al.}

\title [mode = title,
    align=\centering,
    color=black]{Asymptotically Efficient Conformal Inference for Individual Treatment Effects under Multivalued Treatments}                      



\author[1]{Lequn Ge}
\ead{bz987467417@163.com}

\affiliation[1]{organization={School of Statistics and Mathematics},
	addressline={Zhongnan University of Economics and Law}, 
	city={Wuhan},
	state={Hubei},
	postcode={430073},
	country={China}}

\author[2]{Gaorong Li}
\ead{ligaorong@bnu.edu.cn}
\affiliation[2]{organization={School of Statistics},
	addressline={Beijing Normal University}, 
	city={Beijing},
	postcode={100875},
	country={China}}
	
\author[1]{Hui Zhao}[style=chinese]
\ead{hzhao@zuel.edu.cn}
\cormark[1]
\cortext[1]{Corresponding author.}

\begin{abstract}
Most existing conformal inference methods for individual treatment effects are limited to settings with binary treatments. For multiple treatments, propensity score-derived weights often suffer from statistical instability, while contiguous prediction intervals may span low-density regions between modes of potential outcome distributions, leading to inefficient prediction sets. We propose adaptive prediction sets for counterfactual outcomes under multivalued treatments by combining KuLSIF-based density ratio estimation with highest
predictive density (HPD) conformal scores derived from conditional density estimates. Both estimators share a reproducing kernel Hilbert space least-squares formulation, which simplifies theoretical derivation and analysis. The resulting prediction sets are robust to heteroscedasticity and multimodal outcome distributions and effectively eliminate low-density regions between distribution modes. Under unconfoundedness assumption and regularity conditions, we prove that the proposed method converges to optimal HPD sets under covariate shift and establish its asymptotic conditional validity and statistical efficiency. Simulation studies and ACIC benchmarks demonstrate that the proposed method achieves empirical coverage probabilities close to the nominal level, while yielding shorter average interval lengths than conventional quantile-based conformal methods for multimodal outcome data.

\end{abstract}


\maketitle

\section{Introduction}
Causal inference is fundamentally concerned with quantifying the effects of treatments on outcomes~\cite{imbens2015causal}. Early methodological advances in this area primarily focused on causal analysis for binary treatments. However, applied researchers in precision medicine, public policy, and economics routinely select among three or more discrete intervention strategies in their comparative effectiveness research. This growing demand has positioned causal inference with multivalued treatments as a prominent research frontier~\cite{rose2019double,chen2025comparison}.  

Consistent with the binary paradigm, the prevailing literature remains heavily concentrated on population-level average treatment effects (ATE). Such aggregate summaries provide only coarse characterizations of the effect distribution and can be misleading for individualized decision making~\cite{lei2021conformal}. For instance, when comparing three clinical strategies, the drug with the highest average efficacy may prove ineffective or harmful for patients with specific risk profiles. Relative to binary scenarios, multivalued treatments substantially expand the decision space, rendering individual response patterns more complex. Consequently, safe decision making demands a deeper understanding of individual treatment effects (ITE)~\cite{bose2025prediction}.

This necessity has spurred work on conditional average treatment effects (CATE)~\cite{rose2019double,kunzel2019metalearners,wager2018estimation,kim2019estimating}. Nevertheless, CATE characterizes only the conditional mean, and a point estimate falls short of quantifying uncertainty in ITE. Achieving adequate uncertainty quantification through conditioning generally requires a near-perfect CATE estimator and an extremely rich covariate set—conditions rarely satisfied in observational studies. In high-stakes fields such as medicine, there is an urgent need for methods providing finite-sample guarantees. This paper addresses this gap by developing a conformal inference framework for quantifying uncertainty in ITE under multivalued treatments.

Conformal prediction (CP) has emerged as a promising approach in causal inference, offering a flexible, distribution-free framework with rigorous finite-sample coverage guarantees~\cite{lei2021conformal,alaa2023conformal,chen2024conformal}. Lei and Candès~\cite{lei2021conformal} extended the conformalized quantile regression (CQR) framework~\cite{romano2019conformalized} by estimating propensity scores with gradient boosted trees. Recognizing the problem as one of covariate shift, they applied the weighted conformal inference procedure of Tibshirani et al.~\cite{tibshirani2019conformal}, which is based on density ratios. This procedure corrects for the covariate shift induced by treatment assignment in observational data and delivers valid prediction intervals for ITE under binary treatments. Within the weighted conformal framework, Chen et al.~\cite{chen2024conformal} relaxed the unconfoundedness assumption by using neural networks to estimate the density ratio between the experimental and target populations, and consequently constructed intervals for ITE under binary treatments in the target population. To avoid the instability of density ratio weighting under small estimated probabilities and extreme weights, Alaa et al.~\cite{alaa2023conformal} proposed a conformal meta learning framework for binary treatments that uses proxy outcomes for ITE as targets for conformal calibration.

However, existing weighted conformal methods for binary treatments cannot be directly extended to multivalued treatments due to two principal challenges: 
\begin{itemize}
\item \textbf{Extreme weights and effective sample size reduction in multiclass propensity score estimation.} 
Multiclass propensity score estimation for covariate shift correction involves high-dimensional probability modeling with inherent boundary risks. Parametric approaches, such as multinomial logistic regression, introduce misspecification bias in density ratios, whereas nonparametric classifiers like random forests are prone to overfitting, producing extreme weights that reduce effective sample size~\cite{tibshirani2019conformal} and compromise conformal coverage. These shortcomings render propensity-derived density ratios unreliable for multivalued treatment settings, thereby motivating the development of direct density ratio estimation methods.
	
\item \textbf{Structural limitations under complex potential outcome distributions.	}
Existing conformal causal methods predominantly build on CQR~\cite{romano2019conformalized}. While CQR offers finite-sample marginal coverage, asymptotic conditional validity, and some robustness to heteroscedasticity, it produces single contiguous prediction intervals. These intervals often fail to capture multimodality or local gaps in potential outcome distributions, spanning low-density regions and thus becoming unnecessarily wide. Lei and Wasserman~\cite{lei2014distribution} showed that for bounded outcome densities, asymptotic conditional validity follows from asymptotic efficiency, i.e., convergence to the optimal highest predictive density (HPD) set. This stronger notion of efficiency is especially critical under multivalued treatments with complex outcome structures. However, existing HPD-based conformal methods~\cite{izbicki2022cd} rely on an i.i.d. assumption, which precludes their direct application to causal settings with covariate shift.
\end{itemize}

To address these challenges in uncertainty quantification under multivalued treatments, 
we propose a robust weighted conformal framework for constructing compact prediction sets for ITE. We unify density ratio estimation and conformal score construction within a single framework, bypassing multiclass propensity score estimation. Specifically, we use the kernel unconstrained least-squares importance fitting (KuLSIF) method formulated in a reproducing kernel Hilbert space (RKHS)  ~\cite{kanamori2009least,kanamori2012statistical} to directly estimate density ratio weights, yielding stable estimates under complex covariate structures. To construct  more compact and adaptive prediction sets, we employ the cumulative distribution function (CDF) induced by the conditional density as the conformal score~\cite{izbicki2022cd}. This approach approximates HPD regions, reducing excess interval length while accommodating multimodality. Notably, kernel-based conditional density estimation~\cite{sugiyama2010least} shares a mathematical formulation with kernel density ratio estimation, simplifying algorithm design and theoretical analysis.

Finally, we conclude this section by summarizing our contributions and outlining the remainder of this article. The key contributions are threefold:
\begin{itemize}
	\item \textbf{A unified RKHS-based framework for weighted conformal causal inference under multivalued treatments.} We integrate direct density ratio estimation with conformal scores based on the cumulative distribution function induced by an estimated conditional density, thereby eliminating the need to estimate propensity scores. This constitutes a novel framework that constructs adaptive prediction sets for individual treatment effects while providing robust coverage guarantees under multivalued treatments.
	
	\item \textbf{Theoretical guarantees of asymptotic efficiency.} We prove that, under covariate shift induced by multivalued treatment assignment, the proposed weighted method satisfies the stronger property of asymptotic efficiency, thereby providing a theoretical foundation for reliable local inference. 
	
	\item \textbf{Comprehensive empirical validation in complex data settings.} Extensive simulations with heteroscedastic and multimodal outcome distributions, together with evaluations on the ACIC (NSLM) semi-synthetic benchmark, demonstrate that, compared with CQR and classification-based weighting methods, the proposed method maintains stable empirical coverage under extreme distribution shifts and bimodal perturbations while producing more compact prediction sets.
\end{itemize}

The remainder of this article is organized as follows. Section 2 introduces the potential outcome framework for multivalued treatments, formalizes the research objectives, and reviews the relevant foundations of weighted conformal inference. Section 3 presents the proposed robust and adaptive weighted conformal inference framework, describes the construction of compact prediction sets for ITE, and establishes the corresponding theoretical guarantees. Section 4 evaluates the finite-sample performance of the proposed method through simulation studies and experiments on semi-synthetic data. Detailed proofs of all theorems and additional experiments are provided in the Appendix.

\section{Background}

\subsection{Potential Outcomes Framework and Problem Formulation}
Our analysis is based on the classical potential outcomes framework for causal inference~\cite{neyman1923applications,rubin1974estimating}. In contrast to the widely studied binary treatment setting, we focus on discrete multivalued treatments. Let $\mathcal{D} = \{(X_i, Y_i, T_i)\}_{i=1}^n$ denote the observed dataset of $n$ i.i.d. samples. Here, $X_i \in \mathbb{R}^d$ denotes the $d$-dimensional covariate vector for unit $i$; $T_i \in \mathcal{T} = \{1, 2, \dots, K\}$ denotes the treatment received by unit $i$, where $K \geq 2$ is the number of treatment levels; and $Y_i \in \mathbb{R}$ denotes the observed outcome. Under the multivalued treatment framework, each unit $i$ has a potential outcome $Y_i(t)$ for every treatment level $t \in \mathcal{T}$. The fundamental problem of causal inference is that, for any unit $i$, only the single outcome $Y_i(T_i)$ corresponding to the received treatment $T_i$ is observed. The remaining $K-1$ potential outcomes $\{Y_i(t)\}_{t \in \mathcal{T} \setminus \{T_i\}}$ are unobserved and are referred to as counterfactual outcomes. This intrinsic missingness constitutes the fundamental challenge in the inference of individual treatment effects (ITE). To identify treatment effects, we impose the following three standard assumptions.

\textbf{Assumption 1} (Stable Unit Treatment Value Assumption, SUTVA): For any unit $i$ and treatment level $t \in \mathcal{T}$, the potential outcome $Y_i(t)$ does not depend on the treatment assignments of other units (no interference), and consistency holds:  if $T_i = t$, then $Y_i = Y_i(t)$.

\textbf{Assumption 2} (Unconfoundedness): $Y(t) \perp\!\!\!\perp T \mid X, \quad \forall t \in \mathcal{T}$ and all $X$.

\textbf{Assumption 3} (Positivity): $0< P(T=t \mid X) < 1, \quad \forall t \in \mathcal{T}$ and all $X$.

Under Assumptions 2 and 3, for any $a,t \in \mathcal{T}$, we have
\begin{equation}
	P(Y(a) \mid X, T=t) = P(Y(a) \mid X, T=a) = P(Y(a) \mid X).\label{eq:A2}
\end{equation}
Eq.~\eqref{eq:A2} provides the key identity that allows inference on treatment effects under multivalued treatments to be formulated as a covariate shift problem and addressed using weighted conformal prediction.

Given any two treatment levels $a, t \in \mathcal{T} \ (a \neq t)$, most methods for causal inference with multivalued treatments focus on estimating the pairwise CATE~\cite{knaus2022double,acharki2023comparison,zhou2023reference}, defined as
$
\tau_{a,t}(x) = \mathbb{E}[Y(a)-Y(t) \mid X=x].
$
However, CATE point estimates alone do not provide the uncertainty quantification required for high-stakes decision making. We instead focus on inference for ITE, defined for any two treatments $a, t \in \mathcal{T} \ (a \neq t)$ as the contrast $$\tau(a,t) = Y(a) - Y(t).$$ 
Specifically, consider a new test unit with covariates $X_{n+1}$ that received treatment $T_{n+1}=t$. At a nominal coverage level $1-\alpha \in (0,1)$, we construct a prediction interval $\hat{C}_{\mathrm{ITE}}(X_{n+1})$ of the ITE $\tau_{n+1}(a,t)=Y_{n+1}(a) - Y_{n+1}(t)$ for each $a\in \mathcal{T}$ and $a \neq t$. 

Following Lei and Wasserman~\cite{lei2014distribution}, let $P$ be the underlying distribution of a single observation, and $\mathbb{P} = P^{n+1}$ the joint law for the $n$ observed samples and the test point indexed by $n+1$. The target finite-sample guarantee is
$$
\mathbb{P}\left(Y_{n+1}(a)-Y_{n+1}(t) \in \hat{C}_{\text{ITE}}(X_{n+1})\right) \ge 1-\alpha.
$$
Under SUTVA, consistency ensures that $Y_{n+1}(t) = Y_{n+1}$ when the test unit receives treatment $t$, so the outcome $Y_{n+1}(t)$ is observed. Consequently, uncertainty in the ITE $\tau_{n+1}(a,t)$ derives solely from the unobserved potential outcome $Y_{n+1}(a)$. Constructing an ITE interval thus reduces to the construction of a prediction interval $\hat{C}_a(X_{n+1})$ for $Y_{n+1}(a)$ that satisfies
\begin{equation}
	\mathbb{P}\left( Y_{n+1}(a) \in \hat{C}_a(X_{n+1}) \right) \geq 1 - \alpha.\label{eq:mubiao}
\end{equation}
Given $\hat{C}_a(X_{n+1})$, the ITE interval follows by translation: $\hat{C}_{\mathrm{ITE}}(X_{n+1}) = \hat{C}_a(X_{n+1}) - Y_{n+1}$, where $Y_{n+1}$ is the observed factual outcome.

This translation reframes counterfactual inference as a covariate shift problem: we must predict $Y_{n+1}(a)$ for units assigned to treatment $t$ using observed outcomes from treatment group $a$. Because treatment assignment depends on confounders $X$, the covariate distributions differ across groups. Under Assumptions 2 and 3, this misalignment constitutes a covariate shift. 

\subsection{Weighted Conformal Prediction}
To correct the covariate shift in Section 2.1, we adopt weighted conformal prediction (WCP) introduced by Tibshirani et al.~\cite{tibshirani2019conformal}. 
Standard CP constructs prediction intervals without parametric assumptions, providing finite-sample marginal coverage only under exchangeability~\cite{shafer2008tutorial,lei2013distribution,lei2018distribution}. 
In practice, split CP is widely adopted due to its favorable computational efficiency. Specifically, the full dataset is partitioned into a training set $\mathcal{D}_{tr}$ and a calibration set $\mathcal{D}_{ca}$. A model fitted on $\mathcal{D}_{tr}$ yields calibration scores $V_i = V(X_i, Y_i)$ for $i \in \mathcal{D}_{ca}$. Conditional on the training data, the calibration scores and the test score are exchangeable. For a score where larger values indicate greater conformity, the prediction set $\hat{C}(X_{n+1}) = \{y : V(X_{n+1}, y) \ge \hat{q}\}$ satisfies $\mathbb{P}(Y_{n+1} \in \hat{C}(X_{n+1})) \ge 1-\alpha$, where $\hat{q}$ is the adjusted quantile.

However, in our setting, calibration observations (treatment $a$) and the target (treatment $t$) are 
not exchangeable due to covariate misalignment. Under SUTVA and unconfoundedness, Eq.~\eqref{eq:A2} ensures that the conditional distribution of $Y(a)$ given $X$ is invariant across treatment groups, so the discrepancy is purely a covariate shift problem~\cite{shimodaira2000improving}. Assuming absolute continuity, we correct this via the importance weight:
$$w_{t,a}(x) = \frac{\mathrm{d}P_{X|T=t}(x)}{\mathrm{d}P_{X|T=a}(x)}.$$
Estimating this ratio without propensity score pitfalls motivates our approach in Section 3.

\section{Methodology and Theory}
In this section, we develop a practical conformal inference framework for multivalued treatments that avoids the instability of classifier-based propensity scores. Our approach integrates direct density ratio estimation with a flexible conformal score derived from the outcome’s conditional density.

\subsection{Weighted Conformal Causal Inference}
We begin by formalizing the weighted conformal framework for multivalued treatments. Let $\mathcal{D}=\{(X_i,Y_i,T_i)\}_{i=1}^n$ denote the observed data, from which we extract the treatment-specific subsets $\mathcal{D}^{(a)} = \{(X_i, Y_i) \mid T_i = a\}$ and $\mathcal{D}^{(t)} = \{(X_j, Y_j) \mid T_j = t\}$. 

For a new test unit with covariates $X_{n+1}$ that receives treatment $T_{n+1}=t$, we construct a prediction set for the potential outcome $Y_{n+1}(a)$ at the nominal coverage level $1-\alpha$ as follows.

\begin{figure}[pos=htbp]
	\centering
	\includegraphics[width=0.8\textwidth]{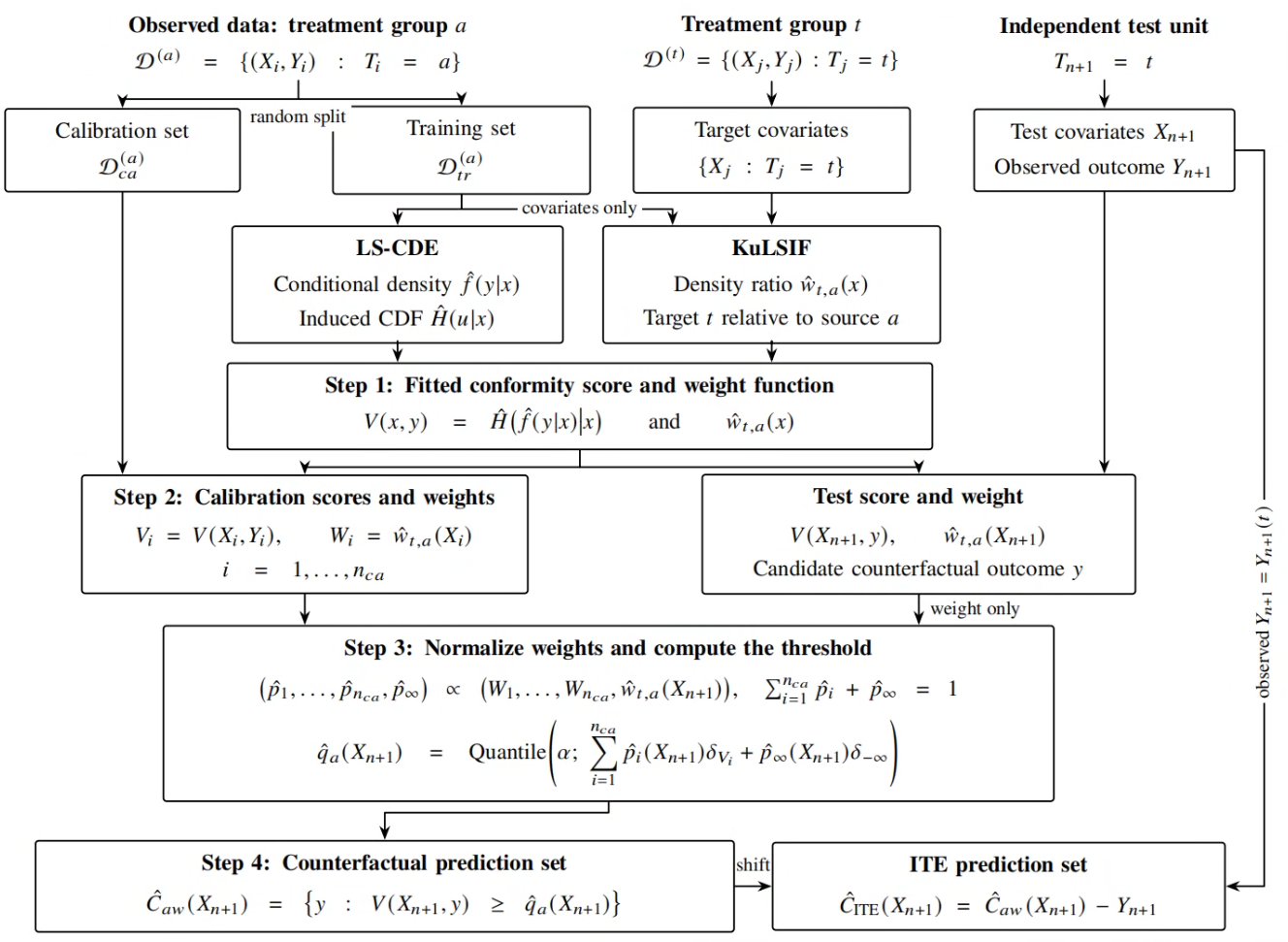}
	\caption{Overview of the Proposed Weighted Conformal Causal Inference Framework}
	\label{fig:4step}
\end{figure}

The proposed method consists of four main steps, with an overview provided in Figure~\ref{fig:4step}.

\begin{itemize}
	
	\item \textbf{Step 1 (Data Splitting and Model Training):} Randomly partition $\mathcal{D}^{(a)}$ into a training set $\mathcal{D}_{tr}^{(a)}$ and a calibration set $\mathcal{D}_{ca}^{(a)}$. We utilize $\mathcal{D}_{tr}^{(a)}$ to (i) estimate the density ratio weight function $\hat{w}_{t,a}(x)$, together with the covariates from $\mathcal{D}^{(t)}$, using KuLSIF as detailed in Section~3.2; and (ii) train the model that underpins the conformal score $V(x,y)$. The score construction follows the HPD principle (Section~3.3) and uses LS-CDE for conditional density estimation (Section~3.4).
	
	\vskip1mm\item \textbf{Step 2 (Computing Calibration Scores):} For each $(X_i,Y_i)\in\mathcal{D}_{ca}^{(a)}$, compute the conformal score $V_i=V(X_i,Y_i)$ using the fitted model from Step 1, and the covariate shift weight $W_i=\hat{w}_{t,a}(X_i)$.
	
	\vskip1mm\item \textbf{Step 3 (Computing the Threshold):} For the test point $X_{n+1}$, define normalized weights for the calibration set and the test point, respectively, as
	\[\begin{aligned}
	    \hat{p}_i(X_{n+1})=&\frac{W_i}{\sum_{j\in\mathcal{D}^{(a)}_{\text{ca}}}W_j+\hat{w}_{t,a}(X_{n+1})},
	\quad i\in\mathcal{D}^{(a)}_{\text{ca}},\\
	\hat{p}_\infty(X_{n+1})=&\frac{\hat{w}_{t,a}(X_{n+1})}{\sum_{j\in\mathcal{D}^{(a)}_{\text{ca}}}W_j+\hat{w}_{t,a}(X_{n+1})}.
	\end{aligned}
	\]
	We then define the threshold $\hat{q}_a(X_{n+1})$ as the weighted $\alpha$-quantile
	\[\begin{aligned}
		\hat{q}_a(X_{n+1})=\text{Quantile}\left(\alpha;\;
		\sum_{i\in\mathcal{D}^{(a)}_{\text{ca}}}\hat{p}_i(X_{n+1})\delta_{V_i}
		+\hat{p}_\infty(X_{n+1})\delta_{-\infty}\right),
	\end{aligned}
	\]
	where $\delta_v$ is a point mass at $v$. The term $\delta_{-\infty}$ safeguards finite-sample coverage by accounting for the unknown test score.
	
	\item \textbf{Step 4 (Constructing the Prediction Set):} For each candidate counterfactual outcome $y \in \mathbb{R}$,we evaluate the score $V(X_{n+1},y)$ and retain all candidates that satisfy $V(X_{n+1},y)\ge\hat{q}_a(X_{n+1})$. The resulting counterfactual prediction set for $Y_{n+1}(a)$ is
	\begin{equation}
		\hat{C}_{aw}(X_{n+1}) = \left\{ y \in \mathbb{R} : V(X_{n+1}, y) \ge \hat{q}_a(X_{n+1}) \right\}.\notag
	\end{equation}
	Since the test unit actually received treatment $t$, we have   
	$Y_{n+1}(t)=Y_{n+1}$, and the weighted conformal ITE prediction set (comparing 
	treatment $a$ to the observed treatment $t$) is the shifted set $$\hat{C}_{\mathrm{ITE}}(X_{n+1}) = \hat{C}_{aw}(X_{n+1}) - Y_{n+1}.$$
\end{itemize}

By Proposition 1 in Lei and Candès~\cite{lei2021conformal}, under SUTVA and unconfoundedness, suppose that $P_{X\mid T=t}$ is absolutely continuous with respect to $P_{X\mid T=a}$, the calibration scores are almost surely distinct, and
$
\left(\mathbb{E}_{X\sim P_{X\mid T=a}}\!\left[\hat{w}_{t,a}(X)^r\right]\right)^{1/r}\le M_r<\infty,r>1.
$
Then the prediction set satisfies
\begin{align}
	1-\alpha-\frac{1}{2}\Delta_w
	&\le
	\mathbb{P}\left(
	Y_{n+1}(a)\in\hat{C}_{aw}(X_{n+1})
	\mid T_{n+1}=t
	\right)\notag\\
	&\le
	1-\alpha+\frac{1}{2}\Delta_w
	+c\,n_{ca}^{1/r-1},
	\label{pro:weighted-coverage}
\end{align}
where
$\Delta_w = \mathbb{E}_{X\sim P_{X\mid T=a}}
\left|\hat{w}_{t,a}(X)-w_{t,a}(X)\right|$, and $c>0$ is a constant depending only on $M_r$ and $r$.
Eq.~\eqref{pro:weighted-coverage} shows that density ratio estimation error controls the potential coverage shortfall, whereas the choice of conformal score primarily governs the size and shape of the prediction set. This distinction motivates the two methodological components presented below.

\subsection{Direct Density Ratio Estimation in an RKHS}
Let $P_{X|T=a}$ and $P_{X|T=t}$ denote the covariate distributions of treatment groups $a$ and $t$, respectively. We assume these distributions admit densities $f_a(x)$ and $f_t(x)$ with respect to a common dominating measure. Following Step 1 in Section 3.1, let $n_a$ and $n_t$ denote the sample sizes of $\mathcal{D}_{tr}^{(a)}$ and $\mathcal{D}^{(t)}$ used for weight estimation. The corresponding covariate samples satisfy
$$X^{(a)}_{1}, \dots, X^{(a)}_{n_a} \overset{i.i.d.}{\sim} P_{X|T=a}, \qquad X^{(t)}_{1}, \dots, X^{(t)}_{n_t} \overset{i.i.d.}{\sim} P_{X|T=t}.$$
Under the absolute continuity condition assumed earlier (cf. Eq.~\eqref{pro:weighted-coverage}), the density ratio of treatment group $t$ with respect to treatment group $a$ is
$$w_{t,a}(x) = \frac{f_t(x)}{f_a(x)},$$
where $w_{t,a}(x)\ge0$. To circumvent error propagation arising from the separate estimation and subsequent ratio computation of $f_t(x)$ and $f_a(x)$, we adopt the kernel unconstrained least-squares importance fitting (KuLSIF) method~\cite{kanamori2009least,kanamori2012statistical} for the direct estimation of $w_{t,a}(x)$.

Let $\hat{w}_{t,a}(x)$ denote an estimator of the density ratio. Consider the population squared error risk with respect to the density $f_a(x)$. Using the identity $w_{t,a}(x)f_a(x)=f_t(x)$, this risk can be expanded as
$$\begin{aligned}
	J(\hat{w}_{t,a}) &= \frac{1}{2}\int (\hat{w}_{t,a}(x) - w_{t,a}(x))^2 f_a(x) \mathrm{d}x  \\
	&= \frac{1}{2}\int \hat{w}^2_{t,a}(x) f_a(x) \mathrm{d}x - \int \hat{w}_{t,a}(x) f_t(x) \mathrm{d}x + C,
\end{aligned}$$
where $C = \frac{1}{2}\int w^2_{t,a}(x) f_a(x) \mathrm{d}x$ is independent of $\hat{w}_{t,a}$ and can thus be ignored during optimization. Suppose that $\hat{w}_{t,a}$ belongs to the RKHS $\mathcal{H}$ . Replacing the expectations with empirical averages and adding a squared RKHS norm penalty to enhance numerical stability and control model complexity yields the regularized empirical objective
$$\widehat{J}(\hat{w}_{t,a}) = \frac{1}{2n_a}\sum_{j=1}^{n_a} \hat{w}^2_{t,a}(x^{(a)}_j)- \frac{1}{n_t}\sum_{l=1}^{n_t} \hat{w}_{t,a}(x^{(t)}_l) + \frac{\lambda}{2}\|\hat{w}_{t,a}\|^2_{\mathcal{H}},$$
where $\lambda > 0$ is the regularization parameter. By the representer theorem, a minimizer admits a finite-dimensional representation
$\hat{w}_{t,a}(x) = \sum_{j=1}^{n_a} \kappa_j k\left(x, x^{(a)}_j\right) + \sum_{l=1}^{n_t} \beta_l k\left(x, x^{(t)}_l\right)$, where $k(\cdot, \cdot)$ is some kernel function, e.g. the Gaussian kernel $k(x, x') = \exp\big(-\frac{\|x-x'\|^2}{2\sigma^2}\big)$ with a kernel bandwidth  $\sigma$. 

By Theorem 1 in Kanamori et al.~\cite{kanamori2012statistical}, the KuLSIF solution admits a representation with $\beta_l=1/(n_t\lambda)$ for $l=1,\dots,n_t$. This structural simplification leaves only $\boldsymbol{\kappa}=(\kappa_1,\dots,\kappa_{n_a})^\top$ to be determined and yields the following representation
\begin{equation*}
	\hat{w}_{t,a}(x) = \sum_{j=1}^{n_a} \kappa_j k\big(x, x_j^{(a)}\big) + \frac{1}{n_t \lambda} \sum_{l=1}^{n_t} k\big(x, x_l^{(t)}\big).
\end{equation*}
%

When $n_a$ is large, $\boldsymbol{\kappa}$ can be obtained by solving the following strictly convex quadratic optimization problem
$$\min_{\boldsymbol{\kappa}} \left\{ \frac{1}{2}\boldsymbol{\kappa}^\top \Big(\frac{1}{n_a}\mathbf{K}_{aa} + \lambda \mathbf{I}_{n_a}\Big)\boldsymbol{\kappa} + \frac{1}{n_a n_t \lambda}\boldsymbol{\kappa}^\top\mathbf{K}_{at}\mathbf{1}_{n_t}  \right\}.$$
Here, $\mathbf{K}_{aa}$ is the Gram matrix with entries $(\mathbf{K}_{aa})_{ij} = k(x^{(a)}_i, x^{(a)}_j)$, $\mathbf{K}_{at}$ is the cross-Gram matrix with $(\mathbf{K}_{at})_{il} = k(x^{(a)}_i, x^{(t)}_l)$, and $\mathbf{1}_{n_t}$ is the $n_t$-dimensional vector of ones. 
Given that the unconstrained estimator may yield negative values, we impose pointwise truncation to derive the KuLSIF estimator
$$\hat{w}^+_{t,a}(x) = \max\{\hat{w}_{t,a}(x), 0\}.$$
The consistency of $\hat{w}^+_{t,a}(x)$ is guaranteed under standard RKHS regularity conditions. We summarize the key result as follows.

\begin{lem}[Theorem 2~\cite{kanamori2012statistical}]
	\label{thm:kuLSIF-asy}
	Let $P_a=P_{X|T=a}$ and suppose that the covariate space $\mathcal{X}$ is compact. Let $\mathcal{H}$ be the RKHS induced by the Gaussian kernel $k(x, x') = \exp\left(-\frac{\|x-x'\|^2}{2\sigma^2}\right)$. Assume that the true density ratio function satisfies $w_{t,a} \in \mathcal{H}$. Let $\mathcal{H}_M$ denote the radius-$M$ ball in $\mathcal{H}$, and suppose that its bracketing entropy under $P_a$ satisfies $H_B(\varepsilon, \mathcal{H}_M, P_a)=O((M/\varepsilon)^\gamma)$ for some constant $\gamma$ such that $0 < \gamma < 2$. Let $\lambda=\lambda_{n_a,n_t}$ satisfy, as $n_a,n_t\to\infty$,
	$$\lambda_{n_a, n_t} \to 0, \quad \lambda_{n_a, n_t}^{-1} = O((n_a \wedge n_t)^{1-\delta}),$$
	where $n_a \wedge n_t = \min\{n_a, n_t\}$, and $\delta$ is a fixed constant satisfying $1 - 2/(2+\gamma) < \delta < 1$. Then the truncated density ratio estimator $\hat{w}^+_{t,a}$ satisfies
	$$\|\hat{w}^+_{t,a} - w_{t,a}\|_{P_a} \le \|\hat{w}_{t,a} - w_{t,a}\|_{P_a} = O_p(\lambda_{n_a, n_t}^{1/2}),$$
	where $\|g\|_{P_a} = \left(\int g^2(x) f_a(x) \mathrm{d}x\right)^{1/2}$.
\end{lem}

The regularization parameter $\lambda$ reflects the bias--variance tradeoff. Lemma~\ref{thm:kuLSIF-asy} implies that, when $\lambda_{n_a,n_t}\to 0$, the density ratio estimation error $\Delta_w$ converges to zero in probability. Consequently, the coverage lower bound in Eq.~\eqref{pro:weighted-coverage} approaches the nominal level $1-\alpha$.

Under the stated regularity conditions, employing a Gaussian kernel yields a convergence rate for the KuLSIF estimator that can be made arbitrarily close to the parametric benchmark  $O_p(1/\sqrt{n_a\wedge n_t})$; see Kanamori et al.~\cite{kanamori2012statistical} for further details.

About the selection of kernel bandwidth $\sigma$ and regularization parameter $\lambda$, we refer to the data-driven leave-one-out cross-validation (LOOCV) squared-error criterion~\cite{kanamori2012statistical}
$$\text{LOOCV} = \frac{1}{n_a \wedge n_t} \sum_{l=1}^{n_a \wedge n_t} \Big\{ \frac{1}{2} \left(\hat{w}^{(-l)+}_{t,a}(x^{(a)}_l)\right)^2 - \hat{w}^{(-l)+}_{t,a}(x^{(t)}_l) \Big\},$$
where $n_a \wedge n_t = \min\{n_a, n_t\}$, and $\hat{w}^{(-l)+}_{t,a}$ denotes the truncated estimator obtained after removing $x^{(a)}_l$ and $x^{(t)}_l$ from their respective treatment groups. The candidate pair $(\sigma, \lambda)$ that minimizes this criterion is selected for the final estimator.

\subsection{The HPD Principle and Conditional CDF as a Conformal Score}

In Step 2 of the procedure described in Section~3.1, the choice of the conformal score $V(X,Y)$ is crucial: it directly affects the shape and compactness of the resulting prediction set. When the conditional potential outcome distribution is multimodal, as it often is in practice, a poorly chosen score may yield a single contiguous interval that spans low-density regions between modes, becoming unnecessarily wide even though it attains the target coverage.


For notational simplicity, we write $f(y|x)$ for the conditional density $f(y(a)|x)=f(y|x, T=a)$ (Assumptions 1-2) of the potential outcome under treatment $a$. For a fixed $x$, among all measurable prediction regions with conditional coverage of at least $1-\alpha$, the set with the smallest Lebesgue measure is, under standard regularity conditions, the highest predictive density (HPD) set~\cite{izbicki2022cd}:
\begin{equation}
	C_\alpha^*(x) := \{y \in \mathbb{R} : f(y|x) > q_\alpha(x)\},\notag
\end{equation}
where $q_\alpha(x)$ is the conditional $\alpha$-quantile of the density value $f(Y|X=x)$. Using the estimated conditional density $\hat{f}(y|x)$ directly as a conformal score is intuitive. However, the conditional distribution of $f(Y|X)$ generally varies with $X$. Consequently, without a localization procedure such as partitioning the covariate space, global calibration based directly on this score does not generally provide asymptotic conditional validity. We therefore adopt HPD-split~\cite{izbicki2022cd} and use the conditional CDF of $f(Y|X)$ as the conformal score. Specifically, define
\begin{equation}
	H(z|x) := \int_{\{y : f(y|x) \le z\}} f(y|x) \mathrm{d}y.\notag
\end{equation}

Given an estimated conditional density $\hat{f}(y|x)$, the corresponding plug-in estimator is
\begin{equation}
	\hat{H}(z|x) = \int_{\{y : \hat{f}(y|x) \le z\}} \hat{f}(y|x) \mathrm{d}y. \notag
\end{equation}

For each calibration observation $(X_i, Y_i)$, the conformal score is then defined as
\begin{equation}
	V_i = \hat{H}(\hat{f}(Y_i|X_i)|X_i),
	\quad i=1,\dots,n_{ca}.\notag
\end{equation}

Since $H(z|x)$ is the conditional CDF of $f(Y|X)$, the random variable $H(f(Y|X)|X)\sim U(0,1)$ given $X$. This conditional uniformity is free of $X$, so the resulting score is approximately independent of $X$ when $\hat{f}(y|x)$ converges to $f(y|x)$. Consequently, the score facilitates the construction of HPD sets that 	
honour the shape of the conditional outcome distribution and adapt to complex scenarios such as multimodality or heteroscedasticity.

Eq.~\eqref{pro:weighted-coverage} establishes finite sample marginal coverage for the counterfactual prediction set. We next show that, under density ratio weighting, HPD-split converges asymptotically to the population optimal HPD set and consequently achieves asymptotic conditional validity~\cite{lei2014distribution}. To this end, we introduce the following standard regularity assumptions, which adapt those used in the unweighted setting~\cite{izbicki2022cd}.

\textbf{Assumption 4:} The observations in the training set $\mathcal{D}_{tr}^{(a)}$ are drawn i.i.d. from the joint distribution induced by $P_{X|T=a} \times P_{Y|X, T=a}$, whereas the new test observation $(X_{n+1}, Y_{n+1}(a))$ is drawn from the joint distribution induced by $P_{X|T=t} \times P_{Y(a)|X}$.

\textbf{Assumption 5:} The covariate distribution $P_{X|T=t}$ is absolutely continuous with respect to  $P_{X|T=a}$. The true density ratio $w_{t,a}(x) = \frac{\mathrm{d} P_{X|T=t}(x)}{\mathrm{d} P_{X|T=a}(x)}\le c$ for a constant $c$ and $P_{X|T=a}$ almost every $x \in \mathcal{X}$. 

\textbf{Assumption 6:} The true conditional density $f(y|x)$ belongs to an RKHS $\mathcal{H}$ induced by a Gaussian kernel on $\mathcal{X} \times \mathcal{Y}$. Moreover, there exists a nonnegative sequence $\eta_n = o(1)$ such that
$$\mathbb{P}\Big(\sup_{x \in \mathcal{X}, y \in \mathcal{Y}} |\hat{f}(y|x) - f(y|x)| \ge \eta_n \Big) = o(1).$$

\textbf{Assumption 7:} For every $x \in \mathcal{X}$, the function $H(u|x)$ is continuous and differentiable with respect to $u$. Moreover, there exist finite constants $M_1$ and $M_2$ such that, in a neighborhood of the theoretical quantile $q_\alpha(x)$, the derivative of $H(u|x)$ is uniformly bounded away from zero and bounded above:
$0 < M_2 \le \frac{\mathrm{d}H(u|x)}{\mathrm{d}u} \le M_1$.

\textbf{Assumption 8:} The covariate space $\mathcal{X}$ and the outcome space $\mathcal{Y}$ are both compact.

Assumption 5 provides the covariate shift adjustment conditions. By Lemma~\ref{thm:kuLSIF-asy}, the density ratio estimator introduced in Section 3.2 yields $\Delta_w = o_p(1)$, ensuring the consistency of the estimated weights. We note that the Gaussian kernel LS-CDE method, detailed in Section 3.4, shares with KuLSIF an RKHS least-squares structure. Furthermore, as established in Lemma~\ref{lem:A.3.1} (see the details in Appendix), the RKHS properties there upgrade the conditional density's $L_2$ convergence (Eq.~\eqref{thm:LSCDE-asy}) to $L_\infty$ uniform convergence, thereby verifying the condition required by Assumption 6 (as shown later).Assumption 7 controls the behavior of the distribution of density values near the HPD threshold: the lower derivative bound ensures local stability of quantile estimation, whereas the upper bound controls the probability mass near the threshold. Assumption 8 restricts the analysis to compact covariate and outcome spaces, ensuring that $\mu_Y(\mathcal{Y})<\infty$ and supporting the Gaussian RKHS entropy and smoothness arguments used in the analysis.

Under these assumptions, the proposed method has the following asymptotic efficiency, which constitutes our main theoretical result.
\begin{thm}[Asymptotic Efficiency]
	\label{thm:asymptotic-hpd}
	Under Assumptions 4--8, the prediction set $\hat{C}_{aw}(X_{n+1})$ constructed using the density ratio weighted HPD conformal procedure is asymptotically equivalent to the optimal highest predictive density set $C_\alpha^*(X_{n+1})$ for treatment group $t$, in the sense that
	$$\mathbb{P}\left(Y_{n+1}(a) \in C_\alpha^*(X_{n+1})\, \Delta\, \hat{C}_{aw}(X_{n+1}) \mid T_{n+1}=t \right) = o(1),$$   where $\Delta$ denotes the symmetric set difference.
	Moreover, the prediction set satisfies asymptotic conditional validity within the treatment group $t$:
	$$\mathbb{P}\left(Y_{n+1}(a) \in \hat{C}_{aw}(X_{n+1}) \mid X_{n+1}, T_{n+1}=t \right) = 1 - \alpha + o_p(1).$$
\end{thm}

A detailed proof is provided in Appendix~\ref{app:proof3}. Existing asymptotic efficiency results for HPD conformal prediction are developed under i.i.d. sampling~\cite{izbicki2022cd}, whereas Theorem~\ref{thm:asymptotic-hpd} extends this property to settings with covariate shift. This result is stronger than the asymptotic conditional validity established for the CQR-based conformal causal framework of Lei and Candès~\cite{lei2021conformal}. Because CQR produces a single contiguous interval, it need not converge to the optimal HPD set when the potential outcome distribution is multimodal or highly skewed. Under the regularity conditions considered here, asymptotic efficiency also implies asymptotic conditional validity~\cite{lei2014distribution}; hence, the proposed method can exclude low-density regions and produce more compact prediction sets while retaining asymptotic conditional validity.

\subsection{Estimating the Conditional Density and Its Induced CDF}

Constructing weighted HPD prediction sets requires estimating the conditional density $f(y| x)$ and the induced CDF $H(u| x)$. A naive approach that estimates the joint and marginal densities separately can amplify estimation errors through their ratio. Instead, we apply least-squares conditional density estimation (LS-CDE)~\cite{sugiyama2010least} to $\mathcal{D}_{tr}^{(a)}$. Notably, LS-CDE and the direct density ratio estimator in Section 3.2 share a common RKHS least-squares structure, which yields analogous convergence guarantees when Gaussian kernels are employed.

Specifically, let $f_a(x)$ denote the density of the covariates in treatment group $a$. The LS-CDE $\hat{f}(y|x)$  minimizes the integrated squared error risk 
	$\iint \big(\hat{f}(y|x) - f(y|x)\big)^2 f_a(x) \mathrm{d}x \mathrm{d}y,$
	where the integration is with respect to the product measure $P_{X|T=a} \times \mu_Y$, and $\mu_Y$ is the Lebesgue measure on $\mathcal{Y}$. 
	
Using the training set $\mathcal{D}_{tr}^{(a)}=\{(x_i^{(a)}, y_i^{(a)})\}_{i=1}^{n_a}$ to approximate the expectations, ignoring terms independent of $\hat{f}(y|x)$, and adding an RKHS norm penalty yields the empirical  object
	\begin{equation}
		\widehat{J}(\hat{f}) = \frac{1}{2n_a} \sum_{i=1}^{n_a} \int \hat{f}(y|x_i^{(a)})^2 \mathrm{d}y - \frac{1}{n_a} \sum_{i=1}^{n_a} \hat{f}(y_i^{(a)}|x_i^{(a)}) + \frac{\lambda'}{2} \|\hat{f}\|_{\mathcal{H}}^2,\label{eq:condition density emp}
	\end{equation}
	where $\lambda' > 0$ is the regularization parameter.
	
To capture nonlinear dependence and multimodality, we employ a product Gaussian kernel basis $\boldsymbol{\phi}(x, y)$ centered at the training observations. The estimator is represented as $\hat{f}(y|x) = \boldsymbol{\theta}^\top \boldsymbol{\phi}(x, y)$, with the $l$-th basis function
	$$\phi_l(x, y) = \exp\Big(-\frac{\|x - x_l^{(a)}\|^2}{2\sigma^2_x}\Big) \exp\Big(-\frac{\|y - y_l^{(a)}\|^2}{2\sigma^2_y}\Big).$$
	Substituting into Eq.~\eqref{eq:condition density emp} gives the convex quadratic objective
	$$\widehat{J}(\boldsymbol{\theta}) = \frac{1}{2} \boldsymbol{\theta}^\top \widehat{\mathbf{H}} \boldsymbol{\theta} - \widehat{\mathbf{h}}^\top \boldsymbol{\theta} + \frac{\lambda'}{2} \boldsymbol{\theta}^\top \mathbf{R} \boldsymbol{\theta},$$
	where $\widehat{\mathbf{H}}= \frac{1}{n_a} \sum_{i=1}^{n_a} \int \boldsymbol{\phi}(x_i^{(a)}, y)\boldsymbol{\phi}(x_i^{(a)}, y)^\top \mathrm{d}y$ is an $n_a \times n_a$ matrix with the $(j,l)$-entry 
	\begin{align}
		\widehat{H}_{jl}
		= &\frac{1}{n_a}\sum_{i=1}^{n_a}
		\exp\!\left(-\frac{\|x_i^{(a)}-x_j^{(a)}\|^2+\|x_i^{(a)}-x_l^{(a)}\|^2}{2\sigma_x^2}\right)\times
		\,(\sqrt{\pi}\sigma_y)^{d_y}\,
		\exp\!\left(-\frac{\|y_j^{(a)}-y_l^{(a)}\|^2}{4\sigma_y^2}\right).\label{eq:A.4}
	\end{align}
	Here $d_y$ is the dimension of the outcome, $\widehat{\mathbf{h}} = \frac{1}{n_a} \sum_{i=1}^{n_a} \boldsymbol{\phi}(x_i^{(a)}, y_i^{(a)})$,
	and $\mathbf{R}$ is the Gram matrix induced by the product kernel basis, representing the RKHS regularization. 
	
	Appendix~\ref{app:proof4} provides the derivation. Since $\lambda' > 0$ and $\mathbf{R}$ is positive definite, $\widehat{\mathbf{H}} + \lambda' \mathbf{R}$ is guaranteed to be nonsingular, and thus the global minimizer is
	$$\tilde{\boldsymbol{\theta}} = \left( \widehat{\mathbf{H}} + \lambda' \mathbf{R} \right)^{-1} \widehat{\mathbf{h}}.$$
	For selecting the kernel bandwidths $\sigma_x$ and $\sigma_y$ and the regularization parameter $\lambda'$, KuLSIF admits the efficient closed form LOOCV criterion in Section 3.2. The LS‑CDE estimator, however, requires additional processing to yield a valid conditional density, which precludes a direct analogue. 
	
	The original LS-CDE enforces $\boldsymbol{\theta} \ge \mathbf{0}$ to avoid negativity~\cite{sugiyama2010least}, but this constrained formulation restricts the solution space and may introduce bias into the fitted density. 
	To preserve the unconstrained least-squares fit $\hat f(y\mid x)=\tilde{\boldsymbol{\theta}}^{\!\top}\boldsymbol{\phi}(x,y)$, we instead apply pointwise truncation and normalization:
	$$\begin{aligned}
		\hat f^+(y\mid x)=&\max\!\bigl(0,\hat f(y\mid x)\bigr),\\
		\tilde f(y_j\mid x)=&\frac{\hat f^+(y_j\mid x)}{\sum_{k=1}^m\hat f^+(y_k\mid x)\,\Delta y},\quad j=1,\dots,m,
	\end{aligned}$$
	where $\mathcal Y_{\text{grid}}=\{y_1,\dots,y_m\}$ is an equally spaced grid with step size $\Delta y$. The normalization is well defined whenever the denominator is strictly positive. 
	Although this post-processing retains the closed‑form unconstrained solution, it breaks the quadratic structure required for the closed form LOOCV in Section 3.2; consequently, hyperparameter tuning must be performed by other means, e.g., 10-fold cross-validation.
	
	This normalization ensures that $\sum_{j=1}^m \tilde{f}(y_j|x) \Delta y = 1$ while preserving the relative shape of the positive part of the unconstrained estimate. The grid representation also enables efficient computation for weighted HPD conformal prediction. In HPD-split, the conformal score for a candidate grid point $y$ is the total probability mass of all regions where the conditional density does not exceed $\tilde{f}(y|x)$. On the discrete grid, this amounts to sorting the grid points by increasing density and accumulating the corresponding probability masses $\tilde{f}(y_j|x)\Delta y$, thereby approximating the conditional cumulative distribution function $H(u|x)$. This approach facilitates efficient construction of weighted conformal prediction sets under covariate shift.
	
	For the subsequent theoretical analysis, we invoke the convergence result established in Theorem 1 of Sugiyama et al.~\cite{sugiyama2010least}, which is the conditional density counterpart of Lemma~\ref{thm:kuLSIF-asy}: 
	Under the regularity conditions of their Theorem 1, suppose that, as $n_a\to\infty$,
	$\lambda'\to0$, and $\lambda'^{-1}=O(n_a^{1-\delta})$,
	where $0<\gamma<2$ is the exponent in the bracketing entropy condition and $\delta$ is a fixed constant satisfying $1-2/(2+\gamma)<\delta<1$. Then the estimation error with respect to the product measure $P_{X|T=a}\times\mu_Y$ satisfies
	\begin{align}
		\|\hat{f}^+(y|x) - f(y|x)\|_{P_{X|T=a} \times \mu_Y}\le \|\hat{f}(y|x) - f(y|x)\|_{P_{X|T=a} \times \mu_Y} = O_p(\lambda'^{1/2}).
		\label{thm:LSCDE-asy}
	\end{align}

\section{Numerical Experiments}
In this section, we evaluate the performance of the proposed method via simulation studies and semi-synthetic experiments. Motivated by Eq.~\eqref{pro:weighted-coverage}, which bounds the deviation of marginal coverage from the nominal level in terms of density ratio estimation error and a finite-sample remainder, our simulation studies examine weight estimation and score construction in two stages. First, holding the conformal score fixed, we compare the empirical marginal coverage obtained with different density ratio estimators across covariate dependence and noise settings. We then use KuLSIF weights for all methods and compare the empirical coverage and prediction set length obtained with different conformal scores across sample sizes. The second comparison focuses on a heteroscedastic bimodal outcome distribution to assess whether HPD-based scores produce more compact prediction sets while maintaining coverage near the nominal level.

\subsection{Simulation Studies}

\subsubsection{Comparison of Marginal Coverage Across Density Ratio Estimators}

We consider the following data generating process:

\begin{itemize}
	\item[] Let $X = (X_1,\dots,X_5)^\top$ with $X_j = \Phi(X_j')$, where $\Phi$ is the standard normal CDF and $(X_1',\dots,X_5') \sim N_5(0,\Sigma)$. The covariance matrix $\Sigma$ has unit diagonal and off-diagonal entries $\rho$. When $\rho = 0$, the components $X_1,\dots,X_5$ are independent standard uniform random variables.
	\item[] Consider the generation process of potential outcomes under three treatments:
	$$\begin{aligned}
		Y(1)=&\frac{2}{1+\exp(-12\cdot(X_1-0.5))}\times\frac{2}{1+\exp(-12\cdot(X_2-0.5))}+\epsilon,\\
		Y(2)=&(X_1-0.5)^2(X_1+1)+(X_2-0.5)^2(X_2+1)+\epsilon,\\
		Y(3) =&\sin(2\pi X_1) + 2|X_2 - 0.5| + \epsilon.
	\end{aligned}$$
	\item[] Let $r^2 = (X_1 - 0.5)^2 + (X_2 - 0.5)^2$. Consider the following noise specifications:(1) Homoscedastic noise: $~\epsilon\sim \mathcal{N}(0,1)$; (2) Heteroscedastic noise: $~\epsilon \sim \mathcal{N}(0, (0.3 + 2.5 r^2)^2)$; (3) Heteroscedastic and bimodal noise:$\epsilon \sim \frac{1}{2}\mathcal{N}(-8r^2, (0.3 + 2.5 r^2)^2) + \frac{1}{2}\mathcal{N}(8r^2, (0.3 + 2.5 r^2)^2).$
	\item[] Observational treatment assignment: The treatment indicators follow a multinomial distribution
	$$(A_1, A_2, A_3) \sim \text{Multinomial}\{e_1(X), e_2(X), e_3(X)\},$$
	where the propensity scores are given by
	$e_k(X) = \frac{\exp(\eta_k(X))}{\sum_{j=1}^3 \exp(\eta_j(X))},$
	with $\tilde{\eta}_1(X) = 0.6 - 5r^2,\, \tilde{\eta}_2(X) = 5 r^2 - 0.6,\,	\tilde{\eta}_3(X) = -2.$
	To achieve roughly balanced groups (each about one-third of the sample) and promote covariate overlap, we center the raw scores,	
	$\eta_k(X) = \tilde{\eta}_k(X) - \text{mean}(\tilde{\eta}_k(X)).$
\end{itemize}

\begin{figure}[pos=htbp]
	\centering
	\includegraphics[width=0.9\textwidth]{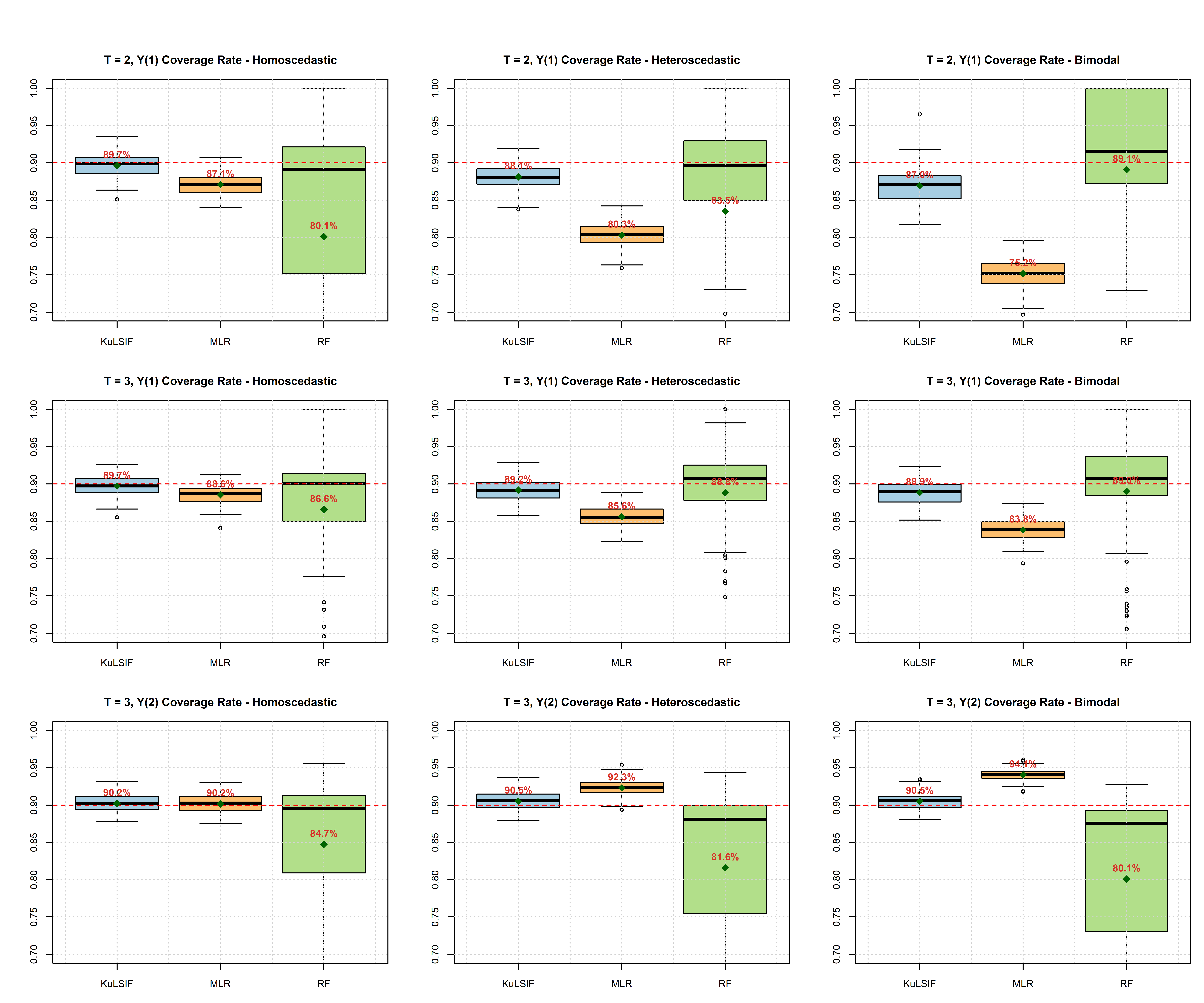}
	\caption{Coverage rates under density ratio weighting with KuLSIF, multinomial logistic regression (MLR), and random forest (RF) for uncorrelated covariates. The red dashed line marks the nominal coverage level of $0.9$, and the green diamonds mark the empirical means. }
	\label{fig:1}
\end{figure}

We parameterize Gaussian kernels as $k_\gamma(u,v)=\exp\{-\gamma\|u-v\|^2\}$, with bandwidth $h=(2\gamma)^{-1/2}$. In each replication, the kernel coefficient and regularization parameter for KuLSIF are selected by leave-one-out cross-validation over $\gamma_{0,K}\in \{0.1,0.5,1,2,5\}, \lambda\in n_{\mathrm{tr}}^{-0.9}\{10^{-5},10^{-4},10^{-3},10^{-2},10^{-1}\}.$For LS-CDE, the kernel coefficient and regularization parameter are jointly selected by ten-fold cross-validation over the candidate grids $\gamma_{0,C}\in\{4.0,4.2,4.4,4.6,4.8,5\}$ and $\lambda\in n_{\mathrm{tr}}^{-0.9}\{10^{-2},10^{-1.5},10^{-1},10^{-0.5},1,10^{0.5},10\}$. The reference coefficients $\gamma_{0,K}$ and $\gamma_{0,C}$ are obtained using the median-distance heuristic implemented in the \texttt{sigest} function of the R package \texttt{kernlab}, and the corresponding multipliers are selected from the candidate grids by cross-validation.

The simulation study examines six scenarios from a $2 \times 3$ factorial design, crossing two covariate dependence settings ($\rho = 0$ or $0.8$) with three noise specifications. Across these scenarios, we compare the empirical coverage of three density ratio estimation methods: RKHS-based direct density ratio estimation, density ratios derived from multinomial logistic regression probabilities, and those derived from random forest (RF) classification probabilities.

For each scenario, we conduct 100 independent replications. In each replication, we generate 6,000 observations and split them evenly into training and calibration sets. The density ratio and conditional density models are fit on the training set. Additionally, an independent test set of 6,000 observations is generated, and prediction sets with nominal 90\% coverage are constructed for the test observations using the methods described above.

Since the treatment groups enter the simulation symmetrically, we evaluate three representative counterfactual prediction tasks: coverage for $Y(1)$ among units assigned to treatment 2, coverage for $Y(1)$ among units assigned to treatment 3, and coverage for $Y(2)$ among units assigned to treatment 3.

Figure~\ref{fig:1} compares the empirical coverage of the three methods under uncorrelated covariates across the three noise settings. For RF probability estimation, the number of candidate features at each split is set to the conventional value $\sqrt{p}$, with 100 trees and a minimum terminal node size of 20. Because only $X_1$ and $X_2$ are active confounders, the RF-based density ratio estimator exhibits substantial instability across the simulation settings, indicating overfitting. The resulting extreme weights reduce the effective sample size and produce considerable variation in coverage across replications. The multinomial logistic regression based estimator is more stable. Under homoscedastic noise, its average coverage is only slightly lower than that achieved by the RKHS-based direct density ratio estimator. Under the two more complex noise settings, however, its coverage is substantially lower, suggesting that the parametric treatment model does not adequately capture the relevant distribution shifts. Overall, among the methods considered, the RKHS based direct density ratio estimator most consistently maintains the nominal coverage level of 90\% across all simulation settings.

\begin{figure}[pos=htbp]
	\centering
	\includegraphics[width=0.9\textwidth]{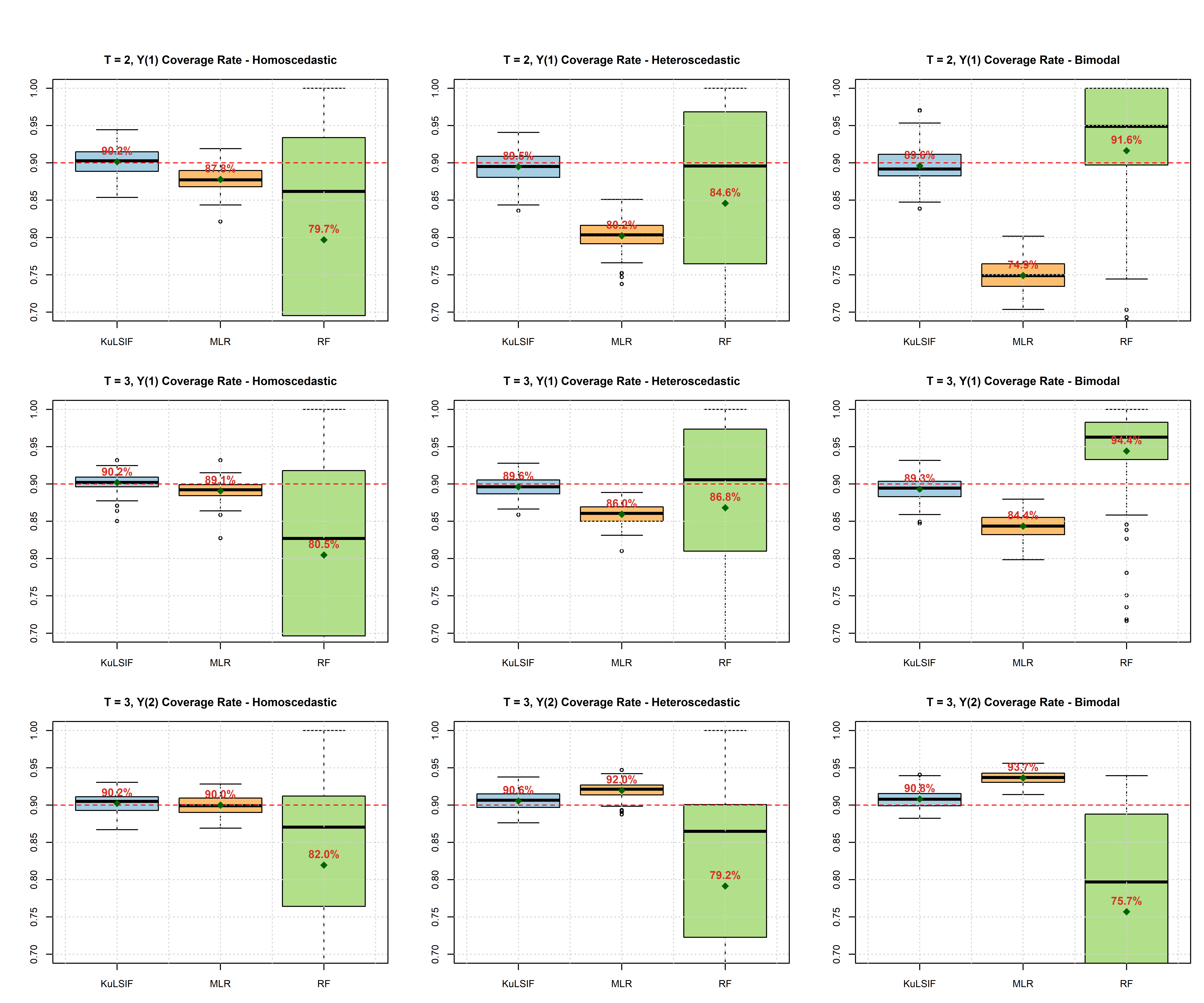}
	\caption{Coverage rates under density ratio weighting with KuLSIF, multinomial logistic regression (MLR), and random forest (RF) for a covariate correlation of $0.8$. The red dashed line marks the nominal coverage level of $0.9$, and the diamonds with adjacent red labels mark the empirical means.  }
	\label{fig:2}
\end{figure}

Figure~\ref{fig:2} compares the empirical coverage across the three noise settings when the pairwise correlation among the covariates is increased to $0.8$. The RF hyperparameters are identical to those used in the uncorrelated setting. Consistent with the preceding results, the RF-based density ratio estimator exhibits the lowest and most variable coverage across the simulation settings, indicating substantial overfitting. Except under homoscedastic noise, the multinomial logistic regression based estimator also yields coverage below the nominal level, suggesting that the parametric treatment model does not adequately capture the distribution shifts under correlated covariates. Among the methods considered, the RKHS-based direct density ratio estimator again maintains the nominal coverage level of 90\% most consistently across all settings.

\subsubsection{Comparison with Alternative Conformal Methods}

The experiments in Section 4.1.1 show that weighting method based on KuLSIF direct density ratio estimation provides the most stable empirical marginal coverage under complex covariate shifts. To further evaluate the relative performance and prediction set efficiency of the proposed weighted HPD method, we uniformly adopt KuLSIF for weight estimation and compare it with commonly used weighted conformal inference methods.

We evaluate the weighted HPD prediction sets using the data generation setting from Section 4.1: constructing prediction sets for the potential outcome $Y(1)$ among units assigned to treatment 2 with highly correlated covariates ($\rho = 0.8$). To highlight differences among conformal inference methods under multimodal outcome distributions, we increase the mode separation coefficient of the heteroscedastic bimodal error term $\epsilon$ to $15$:
$$\epsilon \sim \frac{1}{2}\mathcal{N}\left(-15r^2, (0.3 + 2.5r^2)^2\right) + \frac{1}{2}\mathcal{N}\left(15r^2, (0.3 + 2.5r^2)^2\right).$$
Under this mechanism, as $r^2$ increases and covariate values move toward the boundary of the covariate space, the variance of the error distribution grows sharply, and the distance between the centers of the two Gaussian components also widens. 
This creates a pronounced low-density region between the two modes in the conditional distribution of the response near the boundary. This design evaluates whether each method can exclude the low-density central region to adapt to such distributional structure and produce more compact prediction sets.

In this setting, we compare the weighted HPD-split method with weighted versions of two standard conformal inference approaches:
\begin{itemize}
	\item \textbf{[Local Reg-split]} Normalized split conformal regression based on conformal scores $V_i = |Y_i - \hat{\mu}(x_i)| / \hat{\rho}(x_i)$, where $\hat{\mu}(x)$ estimates the conditional mean and $\hat{\rho}(x_i)$ estimates the conditional mean absolute deviation (MAD), i.e., the conditional mean of $|Y_i-\mu(x_i)|$ given $x_i$~\cite{lei2018distribution}.
	\item \textbf{[Quantile-split]} Split conformal quantile regression~\cite{romano2019conformalized,sesia2020comparison}.
\end{itemize}
For both baseline methods, all regression components are estimated using RFs, including $\hat{\mu}(x)$ and $\hat{\rho}(x)$ in Local Reg-split and the conditional quantiles in split conformal quantile regression.

We conduct 100 independent replicates for sample sizes of 3,000, 6,000, and 9,000. In each replicate, the generated dataset is randomly divided into training and calibration sets of equal size. The training set is used to fit the density ratio model, the conditional density model, and the RF models required by the two baseline methods. Since the data contain three treatment groups and are evenly split between training and calibration, approximately one-sixth of the original observations per treatment group are available for model fitting. Additionally, an independent test set of the same size is generated in each replicate. At a nominal coverage level of $1-\alpha = 90\%$, we compare the mean empirical coverage and mean prediction set length across methods for treatment group $T=2$. Results are reported in Table~\ref{tab:coverage_length}.

\begin{table*}[t]\rmfamily
	\centering
	\renewcommand{\arraystretch}{1}
	\setlength{\tabcolsep}{6pt}
	\caption{\rmfamily Comparison of Average Coverage Rates and Prediction Set Lengths among Conformal Methods Using KuLSIF Weights at Sample Sizes of 3,000, 6,000, and 9,000}
	\label{tab:coverage_length}
	\scalebox{0.85}{
		\resizebox{\textwidth}{!}{
			\begin{tabular}{lcccccc}
				\hline
				\multirow{2}{*}{Method}
				& \multicolumn{2}{c}{$N=3,000$}
				& \multicolumn{2}{c}{$N=6,000$}
				& \multicolumn{2}{c}{$N=9,000$} \\
				\cmidrule(lr){2-3} \cmidrule(lr){4-5} \cmidrule(lr){6-7}
				& Coverage (SD) & Length (SD)
				& Coverage (SD) & Length (SD)
				& Coverage (SD) & Length (SD) \\
				\hline
				Local Reg-split
				& 0.894 (0.027) & 12.30 (0.86)
				& 0.902 (0.021) & 12.25 (0.70)
				& 0.901 (0.017) & 12.06 (0.52) \\
				\addlinespace[0.5ex]
				Quantile-split
				& 0.879 (0.039) & 10.56 (0.84)
				& 0.890 (0.027) & 10.40 (0.52)
				& 0.895 (0.021) & 10.30 (0.39) \\
				\addlinespace[0.5ex]
				HPD-split
				& 0.898 (0.032) & 9.82 (1.10)
				& 0.904 (0.028) & 8.74 (0.75)
				& 0.902 (0.022) & 8.22 (0.53) \\
				\hline
			\end{tabular}
		}
	}
\end{table*}

As shown in Table~\ref{tab:coverage_length}, when KuLSIF weights are used for all methods, the empirical coverage approaches the nominal level of 90\% as the sample size increases, while the corresponding standard deviation across replications decreases. 

Although only about 500 observations per treatment group are available for fitting KuLSIF at the smallest sample size of 3,000, the weighting still adequately corrects for covariate shift and maintains coverage close to the nominal level. Clear differences emerge in prediction set length. 
At each sample size, HPD-split yields the shortest average prediction sets, followed by Quantile-split, while Local Reg-split produces the longest. Moreover, HPD-split shows the greatest reduction in prediction set length as the sample size grows. These results indicate that, under complex heteroscedastic and bimodal distributions, HPD-split constructs more compact prediction sets while preserving reliable empirical coverage.

To illustrate how HPD-split produces more compact prediction sets, we visualize a representative slice along a single covariate, using a dataset of 6,000 observations that yields roughly 1,000 observations per treatment group for model fitting.

\begin{figure}[pos=htbp]
	\centering
	\includegraphics[width=0.9\textwidth]{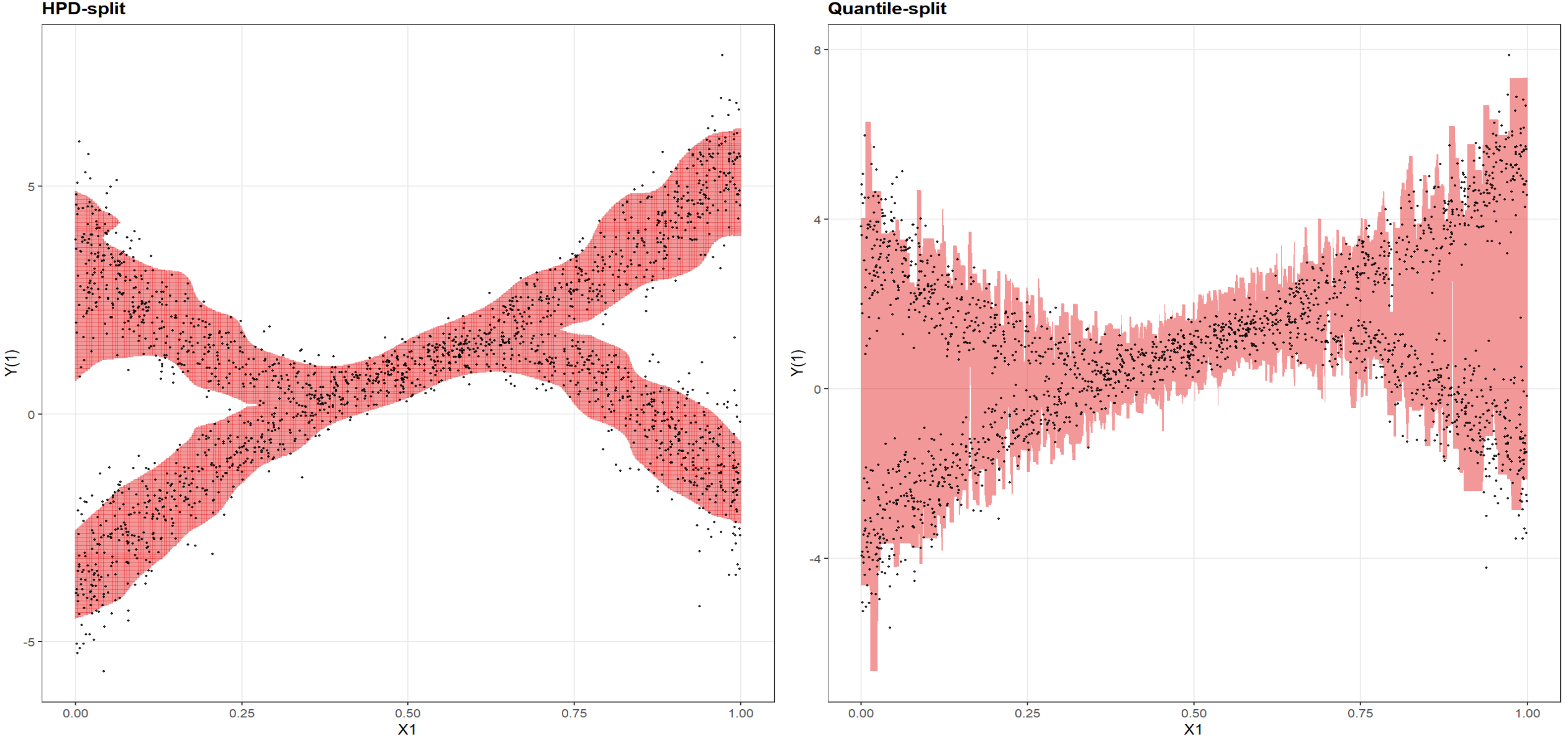}
	\caption{Prediction Set Shapes Along $X_1$ for HPD-split and Quantile-split Scores Under $X_j = 0.5,\ \forall j \neq 1$; Covered Regions in Grey. }
	\label{fig:3}
\end{figure}

We fix all covariates except $X_1$ at the center of their distribution, i.e., set $X_j = 0.5, \forall j \neq 1$, and examine how the shape of the prediction set for the outcome $Y$ varies with $X_1$. The results are shown in Figure~\ref{fig:3}.

\subsection{Performance Evaluation Using Semi-Synthetic Data}

In this section, we evaluate the performance of different conformal inference methods using semi-synthetic data derived from a real study. We draw on data from the National Study of Learning Mindsets (NSLM), previously used in the 2018 Atlantic Causal Inference Conference (ACIC) Data Challenge. Compared to fully synthetic data, these data preserve more realistic nonlinear patterns and complex dependence structures among covariates. Although the original dataset involves a binary treatment, this setting remains relevant, as binary treatment is a special case of multivalued treatment.

The primary objective of this experiment is to construct prediction sets for the potential outcome $Y(1)$ under treatment 1 for the full study population. To simulate the covariate shift and heteroscedasticity commonly encountered in observational studies, we generate semi-synthetic data following the scheme described in Section 4.4 of Lei and Candès~\cite{lei2021conformal}. The procedure is as follows:

First, we randomly partition the original data into a subset $\mathcal{Z}_1$ containing approximately 20\% of the observations and a subset $\mathcal{Z}_2$ containing the remaining 80\%. Base models are fitted using $\mathcal{Z}_1$. A RF estimates the conditional mean in the control group, denoted by $\hat{m}_0(x)$. To capture heteroscedasticity, quantile RFs estimate the 25\% and 75\% conditional quantiles in both treatment groups, from which the conditional interquartile ranges $\hat{r}_0(x)$ and $\hat{r}_1(x)$ are computed. Covariate vectors $X_i$ are then sampled with replacement from $\mathcal{Z}_2$. For each sampled unit, potential outcomes are generated as follows:
\begin{align*}
	Y_i(0) &= \hat{m}_0(X_i) + 0.5 \hat{r}_0(X_i)\epsilon_{i0}, \\
	Y_i(1) &= \hat{m}_0(X_i) + \tau(X_i) + 0.5 \hat{r}_1(X_i)\epsilon_{i1},
\end{align*}
where $\epsilon_{i0}, \epsilon_{i1} \overset{i.i.d.}{\sim} \mathcal{N}(0, 1)$. The function $\tau(x)$ denotes the true CATE, whose specific form is given in Eq. (1) of~\cite{carvalho2019assessing}. Finally, we estimate the propensity score $\hat{e}(x)$ using a RF fitted on $\mathcal{Z}_1$, truncate it to $[0.1, 0.9]$ to enforce the overlap condition, and draw the observed treatment assignment $T_i$ for each sampled unit from a Bernoulli distribution with parameter $\hat{e}(X_i)$.

For sample sizes of 2,500 and 5,000, we conduct 100 independent replications. In each replication, the generated data are equally split into training and calibration sets. On the training set, we fit three density ratio estimators—random forest (RF), logistic regression (LR), and KuLSIF—and then combine each with Local Reg-split, Quantile-split, and HPD-split to construct prediction sets. Based on these sets, we compare the mean empirical coverage and mean prediction set length for the study population. It is important to note that, when constructing prediction sets, the algorithms observe only the covariates $X_i$ of the test observations; the true outcomes $Y_i(1)$ are reserved exclusively for coverage evaluation. As in the previous simulation, all regression components in Local Reg-split and Quantile-split are estimated via RFs, while KuLSIF and HPD-split rely on the kernel estimation procedures described earlier.

\begin{table}[pos=htbp]\rmfamily
	\centering
	\renewcommand{\arraystretch}{1}
	\setlength{\tabcolsep}{6pt}
	\caption{\rmfamily Comparison of Average Coverage Rates and Prediction Set Lengths among Conformal Methods Using Density Ratio Weights Estimated by Random Forest, Logistic Regression, and KuLSIF}
	\label{tab:syn_coverage_length}
	\scalebox{0.85}{
		\resizebox{\textwidth}{!}{
			\begin{tabular}{lcccc}
				\hline
				\multirow{2}{*}{Method Combination}
				& \multicolumn{2}{c}{$N=2,500$}
				& \multicolumn{2}{c}{$N=5,000$} \\
				\cmidrule(lr){2-3} \cmidrule(lr){4-5}
				& Coverage (SD) & Length (SD)
				& Coverage (SD) & Length (SD) \\
				\hline
				LR\_Local Reg-split
				& 0.897 (0.020) & 1.73 (0.13)
				& 0.902 (0.013) & 1.66 (0.08) \\
				\addlinespace[0.5ex]
				LR\_Quantile-split
				& 0.901 (0.021) & 1.63 (0.09)
				& 0.905 (0.014) & 1.58 (0.06) \\
				\addlinespace[0.5ex]
				LR\_HPD-split
				& 0.898 (0.022) & 1.60 (0.10)
				& 0.900 (0.014) & 1.56 (0.06) \\
				\hline
				RF\_Local Reg-split
				& 0.902 (0.021) & 1.76 (0.15)
				& 0.904 (0.014) & 1.67 (0.08) \\
				\addlinespace[0.5ex]
				RF\_Quantile-split
				& 0.905 (0.020) & 1.65 (0.10)
				& 0.905 (0.015) & 1.58 (0.07) \\
				\addlinespace[0.5ex]
				RF\_HPD-split
				& 0.904 (0.022) & 1.63 (0.11)
				& 0.904 (0.014) & 1.58 (0.07) \\
				\hline
				KuLSIF\_Local Reg-split
				& 0.898 (0.020) & 1.73 (0.13)
				& 0.903 (0.013) & 1.66 (0.08) \\
				\addlinespace[0.5ex]
				KuLSIF\_Quantile-split
				& 0.901 (0.020) & 1.63 (0.09)
				& 0.905 (0.014) & 1.58 (0.06) \\
				\addlinespace[0.5ex]
				KuLSIF\_HPD-split
				& 0.899 (0.021) & 1.60 (0.10)
				& 0.902 (0.014) & 1.57 (0.06) \\
				\hline
			\end{tabular}
		}
	}
\end{table}

Table~\ref{tab:syn_coverage_length} summarizes the empirical results on the NSLM semi-synthetic data. First, under the binary treatment setting, density ratio estimation is simpler than under multivalued treatments. The mean empirical coverage of all methods remains close to the nominal level of 0.900, confirming that both KuLSIF weighting and propensity score weights (via LR or RF) adequately correct for covariate shift under the overlap assumption. Second, the methods differ more clearly in prediction set length. At $N = 2,500$, Local Reg-split produces the widest sets (average length 1.73 to 1.76). Quantile-split reduces this to 1.63, reflecting its adaptation to heteroscedasticity; given the symmetric, unimodal Gaussian error distribution from~\cite{lei2021conformal}, Quantile-split is well suited here and achieves asymptotic conditional validity. Nonetheless, across all weighting schemes, HPD-split yields lengths comparable to or slightly shorter than Quantile-split. At $N = 5,000$, standard deviations decrease for all methods, and the relative ordering remains unchanged. These results suggest that HPD-split retains competitive efficiency in this unimodal, heteroscedastic setting.

\section{Conclusion}
Individual treatment effect inference is fundamentally limited by the fact that only one potential outcome is observed per unit. While the existing conformal inference methods primarily focus on binary treatments, extending them to multivalued treatments introduces additional challenges, including the need to estimate multiple density ratios and handle complex covariate shifts across treatment groups. This paper proposes a weighted conformal causal inference framework that systematically addresses these challenges by combining KuLSIF-based weight correction with an HPD conformal score derived from conditional density estimation. The framework constructs valid prediction sets for counterfactual outcomes under multivalued treatments and, when the factual outcome is available, corresponding interval estimates for individual treatment effects. A key methodological contribution is that both components, density ratio estimation and conditional density estimation, share a common least-squares formulation within an RKHS framework, enabling unified computation and asymptotic analysis. Extensive simulations and semi-synthetic experiments demonstrate that weighted HPD-split maintains empirical coverage close to the nominal level while producing substantially more compact prediction sets than existing methods, particularly under complex covariate shift, heteroscedasticity, and multimodality. By providing reliable uncertainty quantification for multivalued treatment settings, the proposed method enhances the practical applicability of conformal causal inference in high-stakes domains such as clinical decision-making and public policy evaluation. Future work may extend the proposed framework to settings with weak overlap or high-dimensional covariates, where density ratio estimation and conditional density estimation become more challenging. Another important direction is relaxing the ignorability assumption to accommodate non-random treatment assignment, though this would require additional structural assumptions.

\vspace{6mm}
\noindent{\bf Acknowledgments}

Zhao's work was supported by the National Natural Science Foundation of China (Grant No. 12571302), and Gaorong Li's work was supported by the National Natural Science Foundation of China (Grant No. 12271046).


\bibliographystyle{cas-model2-names}
\bibliography{ER-Comformal_causal}

\section*{Appendix: Proof of Theorem~\ref{thm:asymptotic-hpd}}
\refstepcounter{section}
\label{app:proof3}

Before presenting the proof of Theorem 1, we first establish several auxiliary lemmas (Lemmas 1-7), which will serve as building blocks for the main argument.

\begin{lem}
	\label{lem:A.3.1}
	Under Assumption 8 of Theorem~\ref{thm:asymptotic-hpd}, which requires $\mathcal{X}$ and $\mathcal{Y}$ to be compact, the Gaussian kernel conditional density estimator described in Section 3.4 satisfies the following uniform convergence property for some nonnegative sequence $\eta_n=o(1)$:
	$$\mathbb{P}\left( \sup_{x \in \mathcal{X}, y \in \mathcal{Y}} |\hat{f}(y|x) - f(y|x)| \ge \eta_n \right) = o(1).$$
\end{lem}

\noindent\textbf{Proof.} 
	Eq.~\eqref{thm:LSCDE-asy} in Section 3.4 gives
	$$\begin{aligned}
		&\|\hat{f}(y|x) - f(y|x)\|_{P_X \times \mu_Y}^2 = \iint (\hat{f}(y|x) - f(y|x))^2 f_a(x) \mathrm{d}x \mathrm{d}y = O_p(\lambda'),
	\end{aligned}
	$$
	where $P_X=P_{X|T=a}$. To convert this weighted $L_2$ bound into a bound under the standard $L_2$ norm, we additionally impose the uniform positivity condition that $f_a(x)$ is bounded away from zero on the compact set $\mathcal{X}$. Specifically, there exists a constant $c_0 > 0$ such that $f_a(x) \ge c_0$ for every $x \in \mathcal{X}$. It follows that
	$$\begin{aligned}
		&\iint \big(\hat{f}(y|x) - f(y|x)\big)^2 \mathrm{d}x \mathrm{d}y \le \frac{1}{c_0} \iint \big(\hat{f}(y|x) - f(y|x)\big)^2 f_a(x) \mathrm{d}x \mathrm{d}y \le \frac{1}{c_0} O_p(\lambda').
	\end{aligned}
	$$
	Therefore, the estimation error converges to zero at the same rate under the standard $L_2$ norm:
	\begin{equation}
		\left( \iint \big(\hat{f}(y|x) - f(y|x)\big)^2 \mathrm{d}x \mathrm{d}y \right)^{1/2} = O_p(\lambda'^{1/2}) = o_p(1). \label{eq:A3.1}
	\end{equation}
	
	The RKHS regularity conditions used to establish Eq.~\eqref{thm:LSCDE-asy} further imply that the regularized conditional density estimator satisfies $\|\hat{f}(y|x)\|_{\mathcal{H}} = O_p(1)$. Moreover, the assumption $f(y|x) \in \mathcal{H}$ in Eq.~\eqref{thm:LSCDE-asy} implies that $\|f(y|x)\|_{\mathcal{H}} < \infty$. The triangle inequality therefore gives
	\begin{equation}
		\|\hat{f}(y|x) - f(y|x)\|_{\mathcal{H}} \le \|\hat{f}(y|x)\|_{\mathcal{H}} + \|f(y|x)\|_{\mathcal{H}} = O_p(1).\label{eq:A3.2}
	\end{equation}
	
	For the interpolation argument, we additionally require $\Omega:=\mathcal{X}\times\mathcal{Y}\subset\mathbb{R}^{d+1}$ to be a bounded Lipschitz domain. The smoothness of the Gaussian kernel implies that the corresponding RKHS $\mathcal{H}$ can be continuously embedded into a Sobolev space of sufficiently high order. In particular, there exists a sufficiently large integer $m$ such that
	$$\mathcal{H} \hookrightarrow H^m(\Omega),$$
	which means that there exists a constant $C_E>0$ such that, for every $f\in\mathcal{H}$,
	$$\|f\|_{H^m(\Omega)} \le C_E \|f\|_{\mathcal{H}}.$$
	Choose a real number $s$ satisfying $\frac{d+1}{2} < s < m$. The Sobolev embedding theorem gives
	$$H^s(\Omega)\hookrightarrow L_\infty(\Omega),$$
	and hence there exists a constant $C_1>0$ such that
	$$\|f\|_{L_\infty(\Omega)} \le C_1 \|f\|_{H^s(\Omega)}.$$
	In addition, an interpolation inequality of Gagliardo--Nirenberg type implies that there exists a constant $C_2>0$ such that
	$$\|f\|_{H^s(\Omega)} \le C_2 \|f\|_{L_2(\Omega)}^{1-s/m} \|f\|_{H^m(\Omega)}^{s/m}.$$
	Combining these inequalities with the continuous embedding of $\mathcal{H}$ into $H^m(\Omega)$ yields
	$$\|f\|_{L_\infty(\Omega)} \le C \|f\|_{L_2(\Omega)}^{1-s/m} \|f\|_{\mathcal{H}}^{s/m},$$
	where $C = C_1 C_2 C_E^{s/m} > 0$. Thus, for any $g_n\in \mathcal{H}$, the supremum norm can be controlled jointly by the $L_2$ norm and the RKHS norm:
	\begin{equation}
		\|g_n\|_{L_\infty(\Omega)} \le C \|g_n\|_{L_2(\Omega)}^{\theta} \|g_n\|_{\mathcal{H}}^{1-\theta},\label{eq:A3.3}
	\end{equation}
	where $\theta = 1-s/m \in (0,1)$.
	
	Let $g_n(x, y) = \hat{f}(y|x) - f(y|x)$. Applying Eq.~\eqref{eq:A3.3} and using Eq.~\eqref{eq:A3.1} and~\eqref{eq:A3.2}, we obtain
	$$\begin{aligned}
		\sup_{x \in \mathcal{X}, y \in \mathcal{Y}} |\hat{f}(y|x) - f(y|x)| &\le C \cdot \left( \|g_n\|_{L_2} \right)^\theta \cdot \left( \|g_n\|_{\mathcal{H}} \right)^{1-\theta} \\
		&= C \cdot \left( O_p(\lambda'^{1/2}) \right)^\theta \cdot \left( O_p(1) \right)^{1-\theta} \\
		&= O_p(\lambda'^{\theta/2}).
	\end{aligned}$$
	Because $\lambda' \to 0$ as $n \to \infty$, we have $\lambda'^{\theta/2} \to 0$. Therefore,
	$$\sup_{x \in \mathcal{X}, y \in \mathcal{Y}} |\hat{f}(y|x) - f(y|x)| = o_p(1).$$
	Equivalently, there exists a deterministic sequence $M_n\to\infty$ that increases sufficiently slowly such that $\eta_n=M_n\lambda'^{\theta/2}=o(1)$ and
	$$\mathbb{P}\left(\sup_{x \in \mathcal{X}, y \in \mathcal{Y}}|\hat{f}(y|x)-f(y|x)|\ge\eta_n\right)=o(1).$$
	This proves the result. $\hfill \square$

\begin{lem}
	\label{lem:A.3.2}
	Under Assumptions 4 - 8 of Theorem~\ref{thm:asymptotic-hpd}, let $X_{n+1}$ be a test covariate drawn from the distribution $P_{X|T=t}$ of treatment group $t$. As the sample size $n \to \infty$, the following statements hold:
	
	{\rm (1)}. $\sup_{u} |H(u|X_{n+1}) - \hat{H}(u|X_{n+1})| = o_p(1)$.
	
	{\rm (2)}. $\hat{H}^{-1}(\alpha|X_{n+1}) = H^{-1}(\alpha|X_{n+1}) + o_p(1)$.
	
\end{lem}

\noindent\textbf{Proof.} 
	We first prove conclusion (1), the consistency of $\hat{H}$. Define the supremum of the conditional density estimation error at the test covariate $X_{n+1}$ as
	$$\epsilon_n = \sup_{y \in \mathcal{Y}} |\hat{f}(y|X_{n+1}) - f(y|X_{n+1})|.$$
	Lemma~\ref{lem:A.3.1} implies that $\epsilon_n = o_p(1)$. Under Assumption 8, the compact set $\mathcal{Y}$ has finite Lebesgue measure, denoted by $\text{Leb}(\mathcal{Y}) < \infty$. For any fixed $u$, define the true and estimated lower level sets, respectively, as $A = \{y \in \mathcal{Y}: f(y|X_{n+1}) \le u\}$ and $\hat{A} = \{y \in \mathcal{Y}: \hat{f}(y|X_{n+1}) \le u\}$. By the definitions of $H$ and $\hat{H}$,
	$$\begin{aligned}
		H(u|X_{n+1}) - \hat{H}(u|X_{n+1}) =\int_A f(y|X_{n+1}) \mathrm{d}y - \int_{\hat{A}} \hat{f}(y|X_{n+1}) \mathrm{d}y.
	\end{aligned}
	$$
	Adding and subtracting $\int_A \hat{f}(y|X_{n+1}) \mathrm{d}y$ gives the decomposition
	$$\begin{aligned}
		H(u|X_{n+1}) - \hat{H}(u|X_{n+1}) = \underbrace{\int_A (f(y|X_{n+1}) - \hat{f}(y|X_{n+1})) \mathrm{d}y}_{\text{Term I}}+ \underbrace{\int_A \hat{f}(y|X_{n+1}) \mathrm{d}y - \int_{\hat{A}} \hat{f}(y|X_{n+1}) \mathrm{d}y}_{\text{Term II}}.
	\end{aligned}
	$$
	Term I satisfies
	$$\begin{aligned}
		\left| \int_A (f(y|X_{n+1}) - \hat{f}(y|X_{n+1})) \mathrm{d}y \right| \le \int_A |f(y|X_{n+1}) - \hat{f}(y|X_{n+1})| \mathrm{d}y 
		\le \epsilon_n \cdot \text{Leb}(A) \le \epsilon_n \cdot \text{Leb}(\mathcal{Y}) = o_p(1).
	\end{aligned}$$
	For Term II, the difference between the integration regions satisfies
	\begin{equation}
		\int_A \hat{f}(y|X_{n+1}) \mathrm{d}y - \int_{\hat{A}} \hat{f}(y|X_{n+1}) \mathrm{d}y \le \int_{A \setminus \hat{A}} \hat{f}(y|X_{n+1}) \mathrm{d}y.\label{eq:A3.4}
	\end{equation}
	For every $y \in A \setminus \hat{A}$, we have $f(y|X_{n+1}) \le u$ and $\hat{f}(y|X_{n+1}) > u$. Hence,
	\begin{equation}
		0 < \hat{f}(y|X_{n+1}) - u \le \hat{f}(y|X_{n+1}) - f(y|X_{n+1}) \le \epsilon_n,\label{eq:A3.5}
	\end{equation}
	which implies that $\hat{f}(y|X_{n+1}) \le u + \epsilon_n$ on $A \setminus \hat{A}$. Substitution into Eq.~\eqref{eq:A3.4} gives
	$$\int_{A \setminus \hat{A}} \hat{f}(y|X_{n+1}) \mathrm{d}y \le (u + \epsilon_n) \cdot \text{Leb}(A \setminus \hat{A}).$$
	Moreover, Eq.~\eqref{eq:A3.5} implies that $u - \epsilon_n < f(y|X_{n+1}) \le u$ on $A \setminus \hat{A}$.
	
	For uniform control over $u$, consider the symmetric difference $A\Delta\hat{A}$. The uniform density error bound gives
	$$A\Delta\hat{A} \subseteq \{y\in\mathcal{Y}:u-\epsilon_n<f(y|X_{n+1})\le u+\epsilon_n\}.$$
	Because $\hat{f}(y|X_{n+1})\le f(y|X_{n+1})+\epsilon_n$, we obtain
	$$\begin{aligned}
		|\text{Term II}|
		&\le \int_{A\Delta\hat{A}}\hat{f}(y|X_{n+1})\mathrm{d}y \\
		&\le H(u+\epsilon_n|X_{n+1})-H(u-\epsilon_n|X_{n+1})
		+\epsilon_n\text{Leb}(\mathcal{Y}).
	\end{aligned}$$
	Under Assumption 7, $H(\cdot|x)$ is a continuous cumulative distribution function and is therefore uniformly continuous in $u$. Since $\epsilon_n=o_p(1)$, 
	$$\sup_u\left\{H(u+\epsilon_n|X_{n+1})-H(u-\epsilon_n|X_{n+1})\right\}=o_p(1).$$
	Combining this result with $\epsilon_n\text{Leb}(\mathcal{Y})=o_p(1)$ and the bound for Term I gives
	$$\sup_{u} |H(u|X_{n+1}) - \hat{H}(u|X_{n+1})| = o_p(1).$$
	
	We next prove conclusion (2), the consistency of the estimated inverse at $\alpha$. Define the inverse as the generalized quantile function, $H^{-1}(\alpha|x)=\inf\{u:H(u|x)\ge\alpha\}$, and define $\hat{H}^{-1}$ analogously. By Assumption 7, $H(\cdot|X_{n+1})$ is continuous and strictly increasing in a neighborhood of $H^{-1}(\alpha|X_{n+1})$. Therefore, for any $\epsilon > 0$ sufficiently small, there exists a $\delta > 0$ such that
	$$H(H^{-1}(\alpha|X_{n+1}) + \epsilon | X_{n+1}) > \alpha + \delta,$$
	and
	$$H(H^{-1}(\alpha|X_{n+1}) - \epsilon | X_{n+1}) < \alpha - \delta.$$
	For a larger value of $\epsilon$, the same argument can be applied using a smaller positive value within this neighborhood.
	
	Define the event
	$$A_n = \left\{ \sup_{u} |\hat{H}(u|X_{n+1}) - H(u|X_{n+1})| < \delta \right\}.$$
	On the event $A_n$,
	\begin{equation}
		\begin{aligned}
			\hat{H}(H^{-1}(\alpha|X_{n+1}) + \epsilon | X_{n+1}) &> H(H^{-1}(\alpha|X_{n+1}) + \epsilon | X_{n+1}) - \delta > \alpha.
		\end{aligned}
		\label{eq:A3.6}
	\end{equation}
	By construction, after truncation and normalization, $\hat{H}(\cdot|X_{n+1})$ is nondecreasing. Eq.~\eqref{eq:A3.6} therefore implies
	\begin{equation}
		\hat{H}^{-1}(\alpha|X_{n+1}) \le H^{-1}(\alpha|X_{n+1}) + \epsilon.\label{eq:A3.7}
	\end{equation}
	Similarly, on the event $A_n$,
	$$\begin{aligned}
		\hat{H}(H^{-1}(\alpha|X_{n+1}) - \epsilon | X_{n+1}) <& H(H^{-1}(\alpha|X_{n+1}) - \epsilon | X_{n+1}) + \delta < \alpha,
	\end{aligned}$$
	which implies
	\begin{equation}
		\hat{H}^{-1}(\alpha|X_{n+1}) \ge H^{-1}(\alpha|X_{n+1}) - \epsilon.\label{eq:A3.8}
	\end{equation}
	
	Conclusion (1) implies that
	$$\lim_{n \to \infty} \mathbb{P}(A_n) = 1.$$
	Combining Eq.~\eqref{eq:A3.7} and~\eqref{eq:A3.8}, we find that the event $A_n$ implies $|\hat{H}^{-1}(\alpha|X_{n+1}) - H^{-1}(\alpha|X_{n+1})| \le \epsilon$. Therefore,
	$$\lim_{n \to \infty} \mathbb{P}\left( |\hat{H}^{-1}(\alpha|X_{n+1}) - H^{-1}(\alpha|X_{n+1})| \le \epsilon \right) = 1.$$
	Since $\epsilon$ is arbitrary, we conclude that
	$$\hat{H}^{-1}(\alpha|X_{n+1}) = H^{-1}(\alpha|X_{n+1}) + o_p(1).$$ $\hfill \square$

\begin{lem}
	\label{lem:A.3.3}
	Under Assumptions 4 - 8 of Theorem~\ref{thm:asymptotic-hpd}, let $X_{n+1}$ be a test covariate drawn from the distribution $P_{X|T=t}$ of treatment group $t$. Define the oracle HPD set as $C_\alpha^*(X_{n+1}) = \{y \in \mathcal{Y} : f(y|X_{n+1}) \ge q_\alpha(X_{n+1})\}$ and the estimated prediction set as $C_\alpha(X_{n+1}) = \{y \in \mathcal{Y} : \hat{f}(y|X_{n+1}) \ge \hat{q}_\alpha(X_{n+1})\}$. Suppose that the conditional density estimator is uniformly consistent and that the threshold estimator is consistent, as established in Lemmas~\ref{lem:A.3.1} and~\ref{lem:A.3.2}; that is, $\sup_{y} |\hat{f}(y|X_{n+1}) - f(y|X_{n+1})| = o_p(1)$ and $|\hat{q}_\alpha(X_{n+1}) - q_\alpha(X_{n+1})| = o_p(1)$. Then, as $n\to\infty$, the conditional probability that the counterfactual outcome $Y_{n+1}(a)$ for a test unit in treatment group $t$ belongs to the symmetric difference between the two sets converges to zero in probability:
	$$\mathbb{P}\left( Y_{n+1}(a) \in C_\alpha^*(X_{n+1}) \Delta C_\alpha(X_{n+1}) \big| X_{n+1} \right) = o_p(1).$$
\end{lem}

\noindent\textbf{Proof.} 
	Throughout the proof, conditioning on the data used to construct $\hat{f}$ and $\hat{q}_\alpha$ is suppressed for notational simplicity. Define the total estimation error
	$$\delta_n = \sup_{y \in \mathcal{Y}} |\hat{f}(y|X_{n+1}) - f(y|X_{n+1})| + |\hat{q}_\alpha(X_{n+1}) - q_\alpha(X_{n+1})|.$$
	By the assumed consistency of the conditional density and threshold estimators, $\delta_n = o_p(1)$. Any $y$ in the symmetric difference $C_\alpha^*(X_{n+1}) \Delta C_\alpha(X_{n+1})$ must belong to one of the following two cases.
	
	Case 1: $y \in C_\alpha(X_{n+1}) \setminus C_\alpha^*(X_{n+1})$.
	
	By the definitions of the sets, $\hat{f}(y|X_{n+1}) \ge \hat{q}_\alpha(X_{n+1})$ and $f(y|X_{n+1}) < q_\alpha(X_{n+1})$. The estimation error bounds give
	$$f(y|X_{n+1}) \ge \hat{f}(y|X_{n+1}) - \delta_n \ge \hat{q}_\alpha(X_{n+1}) - \delta_n \ge q_\alpha(X_{n+1}) - 2\delta_n.$$
	Combining this inequality with $f(y|X_{n+1}) < q_\alpha(X_{n+1})$ yields $q_\alpha(X_{n+1}) - 2\delta_n \le f(y|X_{n+1}) < q_\alpha(X_{n+1})$.
	
	Case 2: $y \in C_\alpha^*(X_{n+1}) \setminus C_\alpha(X_{n+1})$.
	
	In this case, $f(y|X_{n+1}) \ge q_\alpha(X_{n+1})$ and $\hat{f}(y|X_{n+1}) < \hat{q}_\alpha(X_{n+1})$. The same error bounds give
	$$f(y|X_{n+1}) \le \hat{f}(y|X_{n+1}) + \delta_n < \hat{q}_\alpha(X_{n+1}) + \delta_n \le q_\alpha(X_{n+1}) + 2\delta_n.$$
	Combining this inequality with $f(y|X_{n+1}) \ge q_\alpha(X_{n+1})$ yields $q_\alpha(X_{n+1}) \le f(y|X_{n+1}) < q_\alpha(X_{n+1}) + 2\delta_n$.
	
	The two cases imply that every point in the symmetric difference lies within a shrinking band around the true density threshold:
	$$\begin{aligned}
		C_\alpha^*(X_{n+1}) \Delta C_\alpha(X_{n+1}) \subseteq \left\{ y \in \mathcal{Y} : |f(y|X_{n+1}) - q_\alpha(X_{n+1})| \le 2\delta_n \right\}.
	\end{aligned}
	$$
	Therefore, conditionally on the fitted estimators and $X_{n+1}$,
	$$\begin{aligned}
		&\mathbb{P}\left(Y_{n+1}(a) \in C_\alpha^*(X_{n+1}) \Delta C_\alpha(X_{n+1}) \big| X_{n+1} \right) \\
		\le& \mathbb{P}\left( |f(Y_{n+1}(a)|X_{n+1}) - q_\alpha(X_{n+1})| \le 2\delta_n \big| X_{n+1} \right) \\
		=& H\big(q_\alpha(X_{n+1}) + 2\delta_n \big| X_{n+1}\big) - H\big(q_\alpha(X_{n+1}) - 2\delta_n \big| X_{n+1}\big).
	\end{aligned}$$
	The equality follows from the continuity of $H(\cdot|X_{n+1})$ under Assumption 7.
	
	Assumption 7 also states that the derivative of $H(\cdot|X_{n+1})$, which is the density of the random variable $f(Y_{n+1}(a)|X_{n+1})$, is bounded above by $M_1$ in a neighborhood of $q_\alpha(X_{n+1})$. Since $\delta_n=o_p(1)$, the interval $[q_\alpha(X_{n+1})-2\delta_n,q_\alpha(X_{n+1})+2\delta_n]$ lies within this neighborhood with probability tending to one. On this event, the local Lipschitz property of $H$ gives
	$$\begin{aligned}
		H\big(q_\alpha(X_{n+1}) + 2\delta_n \big| X_{n+1}\big) - H\big(q_\alpha(X_{n+1}) - 2\delta_n \big| X_{n+1}\big) \le M_1 \cdot 4\delta_n.
	\end{aligned}
	$$
	Because $M_1<\infty$ and $\delta_n = o_p(1)$, the upper bound $4M_1\delta_n$ converges to zero in probability. The conditional probability is bounded by one outside this event, whose probability converges to zero. Hence,
	$$\mathbb{P}\left( Y_{n+1}(a) \in C_\alpha^*(X_{n+1}) \Delta C_\alpha(X_{n+1}) \big| X_{n+1} \right) = o_p(1).$$ $\hfill \square$

\begin{lem}
	\label{lem:A.3.4}
	Under Assumptions 4 - 8 of Theorem~\ref{thm:asymptotic-hpd}, consider a split asymptotic regime in which the sizes of both the training and calibration sets tend to infinity. Conditional on the training set $\mathcal{D}_{tr}^{(a)}$, let the calibration set $\mathcal{D}_{ca}^{(a)}$ contain $n_{ca}$ observations, and let $X_{n+1}$ denote the test covariate. There exists a positive deterministic sequence $a_{n_{ca}} = o(1)$ such that, for the index set
	$$I_1 = \left\{ i \le n_{ca} : \big| \hat{H}(\hat{f}(Y_i|X_i)|X_i) - H(f(Y_i|X_i)|X_i) \big| \le a_{n_{ca}} \right\},$$
	and its complement $I_2 = \{1, 2, \ldots, n_{ca}\} \setminus I_1$, the following statements hold as $n_{ca} \to \infty$:
	\begin{itemize}
		\item[{\rm (1).}] $|I_1| = n_{ca} + o_p(n_{ca})$, and $|I_2| = o_p(n_{ca})$.
		\item[{\rm (2).}] Define the estimated normalized weights as $\hat{p}_i = \frac{\hat{w}_{t,a}(X_i)}{\sum_{j=1}^{n_{ca}} \hat{w}_{t,a}(X_j) + \hat{w}_{t,a}(X_{n+1})}$. Then the total weight assigned to observations in $I_2$ converges to zero in probability:
		$$\sum_{i \in I_2} \hat{p}_i = o_p(1).$$
	\end{itemize}
\end{lem}

\noindent\textbf{Proof.} 
	We first prove conclusion (1). For any calibration observation $i\le n_{ca}$, the score estimation error satisfies
	$$\sup_{y} |\hat{H}(\hat{f}(y|X_i)|X_i) - H(f(y|X_i)|X_i)| = o_p(1).$$
	To see this, the triangle inequality gives
	$$\begin{aligned}
		\big| \hat{H}(\hat{f}(y|X_i)|X_i) - H(f(y|X_i)|X_i) \big| 
		\le\underbrace{\big| \hat{H}(\hat{f}(y|X_i)|X_i) - H(\hat{f}(y|X_i)|X_i) \big|}_{\text{Term I}}+ \underbrace{\big| H(\hat{f}(y|X_i)|X_i) - H(f(y|X_i)|X_i) \big|}_{\text{Term II}}.
	\end{aligned}$$
	The argument used in Lemma~\ref{lem:A.3.2} applies equally to calibration covariates and shows that Term I converges uniformly to zero in probability. For Term II, Lemma~\ref{lem:A.3.1} gives uniform consistency of $\hat{f}$, while the continuity of $H(\cdot|x)$ in Assumption 7 gives uniform continuity in its first argument. Consequently,
	$$\big| \hat{H}(\hat{f}(y|X_i)|X_i) - H(f(y|X_i)|X_i) \big| = o_p(1).$$
	Therefore, a positive deterministic sequence $a_{n_{ca}}=o(1)$ can be chosen such that, for a generic calibration observation, the event
	$$B_{n_{ca}} := \left\{ \sup_{y} |\hat{H}(\hat{f}(y|X_i)|X_i) - H(f(y|X_i)|X_i)| > a_{n_{ca}} \right\}$$
	satisfies $\mathbb{P}(B_{n_{ca}})=o(1)$. 
	
	Let
	$$p_{n_{ca}}
	=
	\mathbb{P}\left(
	B_{n_{ca}}\mid\mathcal{D}_{tr}^{(a)}
	\right).$$
	Since the score error evaluated at $Y_i$ is bounded by its supremum over $y$, and the calibration observations are identically distributed conditional on the training set, we have
	$$
	\mathbb{E}\left[
	\frac{|I_2|}{n_{ca}}
	\mid\mathcal{D}_{tr}^{(a)}
	\right]
	=
	\frac{1}{n_{ca}}
	\sum_{i=1}^{n_{ca}}
	\mathbb{P}\left(
	i\in I_2\mid\mathcal{D}_{tr}^{(a)}
	\right)
	\le p_{n_{ca}}.
	$$
	Taking expectations yields
	$$
	\mathbb{E}\left[\frac{|I_2|}{n_{ca}}\right]
	\le
	\mathbb{E}[p_{n_{ca}}]
	=
	\mathbb{P}(B_{n_{ca}})
	=
	o(1).
	$$
	Therefore, by Markov's inequality, for every $\varepsilon>0$,
	$$
	\mathbb{P}\left(
	\frac{|I_2|}{n_{ca}}>\varepsilon
	\right)
	\le
	\frac{\mathbb{P}(B_{n_{ca}})}{\varepsilon}
	\to 0.
	$$
	Thus, $|I_2|=o_p(n_{ca})$ and $|I_1|=n_{ca}+o_p(n_{ca})$.
	
	We next prove conclusion (2). The sum of the normalized weights assigned to observations in $I_2$ can be written as
	$$\sum_{i \in I_2} \hat{p}_i = \frac{\frac{1}{n_{ca}} \sum_{i \in I_2} \hat{w}_{t,a}(X_i)}{\frac{1}{n_{ca}} \sum_{j=1}^{n_{ca}} \hat{w}_{t,a}(X_j) + \frac{1}{n_{ca}} \hat{w}_{t,a}(X_{n+1})}.$$
	We first analyze the denominator. By adding and subtracting the true density ratio,
	$$\begin{aligned}
		\left|\frac{1}{n_{ca}}\sum_{j=1}^{n_{ca}}\hat{w}_{t,a}(X_j)-1\right|
		\le \left|\frac{1}{n_{ca}}\sum_{j=1}^{n_{ca}}w_{t,a}(X_j)-1\right| 
		+\frac{1}{n_{ca}}\sum_{j=1}^{n_{ca}}|\hat{w}_{t,a}(X_j)-w_{t,a}(X_j)|.
	\end{aligned}$$
	The first term converges to zero in probability by the law of large numbers because
	$$\begin{aligned}
		\mathbb{E}_{X \sim P_{X|T=a}}[w_{t,a}(X)] = \int_{\mathcal{X}} \frac{\mathrm{d}P_{X|T=t}(x)}{\mathrm{d}P_{X|T=a}(x)} \mathrm{d}P_{X|T=a}(x) = \int_{\mathcal{X}} \mathrm{d}P_{X|T=t}(x) = 1.
	\end{aligned}
	$$
	The conditional expectation of the second term, given the training set, is the $L_1(P_{X|T=a})$ estimation error of $\hat{w}_{t,a}$. Assumption 5 and Markov's inequality therefore imply that this term is $o_p(1)$. Hence,
	$$\frac{1}{n_{ca}}\sum_{j=1}^{n_{ca}}\hat{w}_{t,a}(X_j)=1+o_p(1).$$
	
	Assumption 5 bounds the true density ratio by $w_{\max}$. Moreover, a change of measure gives
	$$\begin{aligned}
		&\mathbb{E}_{X\sim P_{X|T=t}}\left[|\hat{w}_{t,a}(X)-w_{t,a}(X)|\right]\\
		=&\mathbb{E}_{X\sim P_{X|T=a}}\left[w_{t,a}(X)|\hat{w}_{t,a}(X)-w_{t,a}(X)|\right] \\
		\le& w_{\max}\mathbb{E}_{X\sim P_{X|T=a}}\left[|\hat{w}_{t,a}(X)-w_{t,a}(X)|\right]
		=o_p(1).
	\end{aligned}$$
	It follows that $\hat{w}_{t,a}(X_{n+1})=O_p(1)$ and therefore $\frac{1}{n_{ca}}\hat{w}_{t,a}(X_{n+1})=o_p(1)$. Thus, the denominator converges to one in probability.
	
	For the numerator,
	$$\begin{aligned}
		&\frac{1}{n_{ca}} \sum_{i \in I_2} \hat{w}_{t,a}(X_i) \\=& \frac{1}{n_{ca}} \sum_{i=1}^{n_{ca}} \hat{w}_{t,a}(X_i) \mathbb{I}(i \in I_2) \\
		\le& \frac{1}{n_{ca}} \sum_{i=1}^{n_{ca}} w_{t,a}(X_i) \mathbb{I}(i \in I_2) + \frac{1}{n_{ca}} \sum_{i=1}^{n_{ca}} |\hat{w}_{t,a}(X_i) - w_{t,a}(X_i)|.
	\end{aligned}$$
	The second term is the empirical average of the absolute weight estimation errors. Its conditional expectation is the $L_1(P_{X|T=a})$ error in Assumption 5, so it is $o_p(1)$ by Markov's inequality. Since $w_{t,a}(x)\le w_{\max}$ for $P_{X|T=a}$ almost every $x$, the first term satisfies
	$$\frac{1}{n_{ca}} \sum_{i=1}^{n_{ca}} w_{t,a}(X_i) \mathbb{I}(i \in I_2) \le w_{\max} \cdot \frac{|I_2|}{n_{ca}}.$$
	Conclusion (1) gives $|I_2|/n_{ca}=o_p(1)$, and hence the first term is also $o_p(1)$. The numerator therefore converges to zero in probability, whereas the denominator converges to one in probability. It follows that
	$$\sum_{i \in I_2} \hat{p}_i = o_p(1).$$ $\hfill \square$

\begin{lem}
	\label{lem:A.3.5}
	Under Assumptions 4 - 8 of Theorem~\ref{thm:asymptotic-hpd}, let $V_i = \hat{H}(\hat{f}(Y_i|X_i)|X_i)$ and $V_i^* = H(f(Y_i|X_i)|X_i)$ denote the estimated and true positive conformal scores, respectively, for the observations in the calibration set $\mathcal{D}_{ca}^{(a)}$. For a given test covariate $X_{n+1}$ and any quantile level $\alpha \in (0,1)$, let $\hat{q}_{w}(X_{n+1})$ be the $\alpha$ quantile of the modified weighted empirical distribution
	$$\sum_{i \in \mathcal{D}_{ca}^{(a)}} \hat{p}_i(X_{n+1}) \delta_{V_i} + \hat{p}_\infty(X_{n+1}) \delta_{-\infty},$$
	where $\hat{p}_i(X_{n+1})$ and $\hat{p}_\infty(X_{n+1})$ are the normalized weights computed from the estimated density ratio $\hat{w}_{t,a}$, and $\delta_{-\infty}$ denotes a Dirac mass at negative infinity, corresponding to the smallest possible score. Then,
	$$\hat{q}_{w}(X_{n+1}) = \alpha + o_p(1).$$
\end{lem}

\noindent\textbf{Proof.} 
	Retain the index sets $I_1$ and $I_2$ defined in Lemma~\ref{lem:A.3.4}, and let their total weights be $W(I_1) = \sum_{i \in I_1} \hat{p}_i(X_{n+1})$ and $W(I_2) = \sum_{i \in I_2} \hat{p}_i(X_{n+1})$, respectively. On the event $W(I_1)>0$, define the weighted empirical cumulative distribution functions based on the estimated and true scores in $I_1$ as
	\begin{align*}
		\hat{G}_1(t) &= \frac{1}{W(I_1)} \sum_{i \in I_1} \hat{p}_i(X_{n+1}) \mathbb{I}(V_i \le t), \\
		G_1(t) &= \frac{1}{W(I_1)} \sum_{i \in I_1} \hat{p}_i(X_{n+1}) \mathbb{I}(V_i^* \le t).
	\end{align*}
	When $W(I_2)>0$, define the corresponding true weighted empirical distribution on $I_2$ by $$G_2(t) = \frac{1}{W(I_2)} \sum_{i \in I_2} \hat{p}_i(X_{n+1}) \mathbb{I}(V_i^* \le t).$$ 
	If $W(I_2)=0$, define $G_2$ arbitrarily because it is always multiplied by $W(I_2)$. Finally, define the global weighted empirical distribution based on the true scores as
	\begin{equation}
		G_0(t) = \sum_{i \in \mathcal{D}_{ca}^{(a)}} \hat{p}_i(X_{n+1}) \mathbb{I}(V_i^* \le t) + \hat{p}_\infty(X_{n+1}) \mathbb{I}(-\infty \le t). \notag
	\end{equation}
	
	By the definition of $I_1$ in Lemma~\ref{lem:A.3.4}, every $i\in I_1$ satisfies $V_i^* - a_{n_{ca}} \le V_i \le V_i^* + a_{n_{ca}}$. Therefore, for any real threshold $t$, the following event inclusions hold:
	\begin{itemize}
		\item If $\{V_i \le t\}$ occurs, then $V_i^* \le t+a_{n_{ca}}$, and hence $\{V_i \le t\} \subseteq \{V_i^* \le t + a_{n_{ca}}\}$.
		\item If $\{V_i^* \le t-a_{n_{ca}}\}$ occurs, then $V_i \le t$, and hence $\{V_i^* \le t - a_{n_{ca}}\} \subseteq \{V_i \le t\}$.
	\end{itemize}
	These inclusions imply
	\begin{equation}
		\mathbb{I}(V_i^* \le t - a_{n_{ca}}) \le \mathbb{I}(V_i \le t) \le \mathbb{I}(V_i^* \le t + a_{n_{ca}}) .\label{eq:A3.9}
	\end{equation}
	Multiplying Eq.~\eqref{eq:A3.9} by the nonnegative relative weight $\frac{\hat{p}_i(X_{n+1})}{W(I_1)}$ and summing over $i\in I_1$ gives
	\begin{equation}
		G_1(t - a_{n_{ca}}) \le \hat{G}_1(t) \le G_1(t + a_{n_{ca}}).\label{eq:A3.10}
	\end{equation}
	Eq.~\eqref{eq:A3.10} then implies that, for every $\alpha^*\in(0,1)$,
	$$G_1^{-1}(\alpha^*) - a_{n_{ca}} \le \hat{G}_1^{-1}(\alpha^*) \le G_1^{-1}(\alpha^*) + a_{n_{ca}}.$$
	Since $a_{n_{ca}} = o(1)$, it follows that
	\begin{equation}
		\hat{G}_1^{-1}(\alpha^*) = G_1^{-1}(\alpha^*) + o_p(1).\label{eq:A3.11}
	\end{equation}
	
	Let $F_U$ denote the cumulative distribution function of the standard uniform distribution. Under Assumption 7, the conditional probability integral transform gives $V_i^*|X_i\sim\text{Uniform}(0,1)$. In particular, $V_i^*$ is independent of $X_i$. The boundedness of the true density ratio, the $L_1(P_{X|T=a})$ consistency of $\hat{w}_{t,a}$, and the weighted uniform law of large numbers therefore imply
	$$\sup_t|G_0(t)-F_U(t)|=o_p(1).$$
	For every finite $t$,
	$$G_0(t) = W(I_1)G_1(t) + W(I_2)G_2(t) + \hat{p}_\infty(X_{n+1}) \mathbb{I}(-\infty \le t).$$
	Because $0\le G_1(t),G_2(t)\le1$ and $W(I_1)+W(I_2)+\hat{p}_\infty(X_{n+1})=1$, we have
	\begin{equation}
		\sup_t|G_1(t)-G_0(t)|\le W(I_2)+\hat{p}_\infty(X_{n+1})=o_p(1).\label{eq:A3.12}
	\end{equation}
	Here, the final equality follows from Lemma~\ref{lem:A.3.4} and the fact that $\hat{p}_\infty(X_{n+1})=o_p(1)$. Consequently,
	$$\sup_t|G_1(t)-F_U(t)|=o_p(1).$$
	Since $F_U(t)=t$ on $[0,1]$, standard quantile stability under uniform convergence, together with Eq.~\eqref{eq:A3.11}, gives, for every fixed $\eta\in(0,1/2)$,
	$$\sup_{\alpha^*\in[\eta,1-\eta]}|\hat{G}_1^{-1}(\alpha^*)-\alpha^*|=o_p(1).$$
	In particular,
	\begin{equation}
		\hat{G}_1^{-1}(\alpha^*) = \alpha^* + o_p(1).\label{eq:A3.13}
	\end{equation}
	
	Finally, let $\hat{G}_0$ denote the global weighted empirical distribution based on the estimated scores, so that $\hat{q}_{w}(X_{n+1}) = \hat{G}_0^{-1}(\alpha)$. Define $\hat{G}_2$ on $I_2$ analogously to $\hat{G}_1$. For every finite $t$,
	$$\hat{G}_0(t) = W(I_1)\hat{G}_1(t) + W(I_2)\hat{G}_2(t) + \hat{p}_\infty(X_{n+1}).$$
	Since $0\le\hat{G}_2(t)\le1$,
	$$\begin{aligned}
		W(I_1)\hat{G}_1(t) + \hat{p}_\infty(X_{n+1}) \le \hat{G}_0(t) \le W(I_1)\hat{G}_1(t) + W(I_2) + \hat{p}_\infty(X_{n+1}).
	\end{aligned}$$
	On the event that the two arguments below belong to $(0,1)$, inversion of these inequalities gives
	$$\begin{aligned}
		\hat{G}_1^{-1}\left( \frac{\alpha - W(I_2) - \hat{p}_\infty(X_{n+1})}{W(I_1)} \right) \le \hat{q}_{w}(X_{n+1})
		 \le \hat{G}_1^{-1}\left( \frac{\alpha - \hat{p}_\infty(X_{n+1})}{W(I_1)} \right).
	\end{aligned}
	$$
	By Lemma~\ref{lem:A.3.4}, $W(I_2)=o_p(1)$, $\hat{p}_\infty(X_{n+1})=o_p(1)$, and $W(I_1)\xrightarrow{p}1$. Hence, both arguments of $\hat{G}_1^{-1}$ converge in probability to $\alpha$ and belong to a compact subset of $(0,1)$ with probability tending to one. The uniform quantile convergence established above therefore shows that both bounds equal $\alpha+o_p(1)$. The result follows by the squeeze argument:
	$$\hat{q}_{w}(X_{n+1}) = \alpha + o_p(1).$$  $\hfill \square$

\begin{lem}
	\label{lem:A.3.6}
	Let the oracle set $C_\alpha^*(X_{n+1})$ and the estimated prediction set $C_\alpha(X_{n+1})$ be defined as in Lemma~\ref{lem:A.3.3}. All probabilities are evaluated under the distribution of treatment group $t$, with conditioning on $T_{n+1}=t$ suppressed for notational simplicity. By the definition of the oracle set and the continuity condition in Assumption 7,
	$$\mathbb{P}(Y_{n+1}(a) \in C_\alpha^*(X_{n+1}) \mid X_{n+1}) \equiv 1 - \alpha.$$
	Suppose that the symmetric difference between the estimated and oracle sets satisfies
	$$\mathbb{P}\Big(Y_{n+1}(a) \in C_\alpha^*(X_{n+1}) \Delta C_\alpha(X_{n+1})\Big) = o(1).$$
	Then $C_\alpha(X_{n+1})$ satisfies asymptotic conditional validity:
	$$\mathbb{P}(Y_{n+1}(a) \in C_\alpha(X_{n+1}) \mid X_{n+1}) = 1-\alpha+o_p(1).$$
\end{lem}

\noindent\textbf{Proof.} 
	For each $x\in\mathcal{X}$, define
	$$Z_n(x) := \mathbb{P}\Big(Y_{n+1}(a) \in C_\alpha^*(x) \Delta C_\alpha(x) \mid X_{n+1}=x\Big),$$
	and let $Z_n=Z_n(X_{n+1})$. By the law of total expectation,
	$$\mathbb{E}[Z_n] = \mathbb{P}\Big(Y_{n+1}(a) \in C_\alpha^*(X_{n+1}) \Delta C_\alpha(X_{n+1})\Big) = o(1).$$
	Since $Z_n$ is nonnegative and bounded by one, Markov's inequality gives, for every $\varepsilon>0$,
	$$\mathbb{P}(Z_n>\varepsilon)\le\frac{\mathbb{E}[Z_n]}{\varepsilon}\to0.$$
	Thus, $Z_n=o_p(1)$. Consequently, there exists a positive deterministic sequence $\eta_n=o(1)$ such that, for
	$$\Lambda_n=\{x\in\mathcal{X}:Z_n(x)\le\eta_n\},$$
	we have
	\begin{equation}
		\mathbb{P}(X_{n+1} \in \Lambda_n) = 1 - o(1).\label{eq:A3.14}
	\end{equation}
	
	For every $x_{n+1}\in\Lambda_n$,
	$$\begin{aligned}
		&\Big| \mathbb{P}\big(Y_{n+1}(a) \in C_\alpha(x_{n+1}) \mid X_{n+1} = x_{n+1}\big) - (1 - \alpha) \Big| \\
		= {}&\Big| \mathbb{P}\big(Y_{n+1}(a) \in C_\alpha(x_{n+1}) \mid X_{n+1} = x_{n+1}\big)- \mathbb{P}\big(Y_{n+1}(a) \in C_\alpha^*(x_{n+1}) \mid X_{n+1} = x_{n+1}\big) \Big| \\
		\le {}&\mathbb{P}\Big(Y_{n+1}(a) \in C_\alpha^*(x_{n+1}) \Delta C_\alpha(x_{n+1}) \mid X_{n+1} = x_{n+1}\Big) \\
		= {}&Z_n(x_{n+1})\le\eta_n=o(1).
	\end{aligned}$$
	Eq.~\eqref{eq:A3.14} shows that $X_{n+1}\in\Lambda_n$ with probability tending to one. Therefore, the conditional coverage error converges to zero in probability, and
	$$\mathbb{P}(Y_{n+1}(a) \in C_\alpha(X_{n+1}) \mid X_{n+1}) = 1 - \alpha + o_p(1).$$
$\hfill \square$

Combining the preceding lemmas completes the proof of Theorem~\ref{thm:asymptotic-hpd}. Assume that Assumptions 4 - 8 hold. Throughout the proof, all probabilities are evaluated under the distribution of treatment group $t$, with conditioning on $T_{n+1}=t$ suppressed for notational simplicity.

\vskip3mm\noindent{\bf Proof of Theorem 1.}

First, observe that
$$\begin{aligned}
	C_\alpha(X_{n+1}) &= \{y \in \mathcal{Y} : \hat{H}(\hat{f}(y|X_{n+1})|X_{n+1}) \ge \hat{q}_w(X_{n+1})\} \\
	&= \{y \in \mathcal{Y} : \hat{f}(y|X_{n+1}) \ge \hat{H}^{-1}(\hat{q}_w(X_{n+1})|X_{n+1})\}.
\end{aligned}$$
Define the estimated density threshold as $\hat{q}_\alpha(X_{n+1}) = \hat{H}^{-1}(\hat{q}_w(X_{n+1})|X_{n+1})$. Recall that the oracle HPD set is $C_\alpha^*(X_{n+1}) = \{y \in \mathcal{Y} : f(y|X_{n+1}) \ge q_\alpha(X_{n+1})\}$, where $q_\alpha(X_{n+1}) = H^{-1}(\alpha|X_{n+1})$ is the corresponding theoretical density threshold.

Lemma~\ref{lem:A.3.5} gives $\hat{q}_w(X_{n+1}) = \alpha + o_p(1)$. Moreover, the argument underlying Lemma~\ref{lem:A.3.2}, together with the strictly positive lower bound $M_2$ in Assumption 7, gives local uniform consistency of the inverse:
$$\sup_{\beta\in[\alpha-\eta,\alpha+\eta]}\left|\hat{H}^{-1}(\beta|X_{n+1})-H^{-1}(\beta|X_{n+1})\right|=o_p(1)$$
for some $\eta>0$. Assumption 7 also implies that $H^{-1}(\cdot|X_{n+1})$ is locally Lipschitz around $\alpha$. Since $\hat{q}_w(X_{n+1})\in[\alpha-\eta,\alpha+\eta]$ with probability tending to one, these results imply
$$\begin{aligned}
	|\hat{q}_\alpha(X_{n+1}) - q_\alpha(X_{n+1})| = \big| \hat{H}^{-1}(\hat{q}_w(X_{n+1})|X_{n+1}) - H^{-1}(\alpha|X_{n+1}) \big| = o_p(1).
\end{aligned}
$$
Lemma~\ref{lem:A.3.1} further gives
$$\sup_y |\hat{f}(y|X_{n+1}) - f(y|X_{n+1})| = o_p(1).$$

Thus, all conditions of Lemma~\ref{lem:A.3.3} are satisfied. Applying that lemma yields
$$\mathbb{P}\left( Y_{n+1}(a) \in C_\alpha^*(X_{n+1}) \Delta C_\alpha(X_{n+1}) \big| X_{n+1} \right) = o_p(1).$$
This establishes convergence of the estimated prediction set to the oracle HPD set in conditional probability.

To obtain the corresponding marginal statement, define
$$Z_n := \mathbb{P}\left( Y_{n+1}(a) \in C_\alpha^*(X_{n+1}) \Delta C_\alpha(X_{n+1}) \big| X_{n+1} \right).$$
Since $0 \le Z_n \le 1$ and $Z_n = o_p(1)$, for every $\varepsilon > 0$,
$$\mathbb{E}[Z_n] = \mathbb{E}[Z_n \mathbb{I}(Z_n \le \varepsilon)] + \mathbb{E}[Z_n \mathbb{I}(Z_n > \varepsilon)] \le \varepsilon \cdot 1 + 1 \cdot \mathbb{P}(Z_n > \varepsilon).$$
Because $\mathbb{P}(Z_n > \varepsilon) \to 0$ for every fixed $\varepsilon>0$,
$$\limsup_{n \to \infty} \mathbb{E}[Z_n] \le \varepsilon.$$
Letting $\varepsilon \to 0$ gives $\mathbb{E}[Z_n]\to0$. By the law of total expectation,
$$\mathbb{P}\Big( Y_{n+1}(a) \in C_\alpha^*(X_{n+1}) \Delta C_\alpha(X_{n+1}) \Big) = \mathbb{E}[Z_n] = o(1).$$
This proves the asymptotic convergence of the estimated set to the oracle HPD set.

The preceding marginal convergence satisfies the conditions of Lemma~\ref{lem:A.3.6}. Since
$$\mathbb{P}(Y_{n+1}(a) \in C_\alpha^*(X_{n+1}) \mid X_{n+1}) \equiv 1 - \alpha,$$
Lemma~\ref{lem:A.3.6} yields the asymptotic conditional validity of the estimated prediction set:
$$\mathbb{P}\left(Y_{n+1}(a) \in C_\alpha(X_{n+1}) \big| X_{n+1}\right) = 1 - \alpha + o_p(1).$$
$\hfill \square$

\subsection{Derivation of Eq.~\eqref{eq:A.4}}
\label{app:proof4}

The derivation below assumes that integration with respect to $y$ is performed over $\mathbb{R}^{d_y}$. By the definition of the Gaussian basis functions, the product of the basis functions indexed by $j$ and $l$, evaluated at $(x_i^{(a)}, y)$, is
$$\begin{aligned}
	\phi_j(x_i^{(a)}, y) \phi_l(x_i^{(a)}, y) &= \left[ \exp\left(-\frac{\|x_i^{(a)} - x_j^{(a)}\|^2}{2\sigma_x^2}\right) \exp\left(-\frac{\|y - y_j^{(a)}\|^2}{2\sigma_y^2}\right) \right] \\&
	 \times \left[ \exp\left(-\frac{\|x_i^{(a)} - x_l^{(a)}\|^2}{2\sigma_x^2}\right) \exp\left(-\frac{\|y - y_l^{(a)}\|^2}{2\sigma_y^2}\right) \right].
\end{aligned}$$
Substituting this expression into the integral, the terms involving $x$ can be taken outside the integral because they do not depend on $y$:
$$\begin{aligned}
	\int \phi_j(x_i^{(a)}, y) \phi_l(x_i^{(a)}, y) \mathrm{d}y = \exp\left(-\frac{\|x_i^{(a)} - x_j^{(a)}\|^2 + \|x_i^{(a)} - x_l^{(a)}\|^2}{2\sigma_x^2}\right)\times \underbrace{\int \exp\left(-\frac{\|y - y_j^{(a)}\|^2 + \|y - y_l^{(a)}\|^2}{2\sigma_y^2}\right) \mathrm{d}y}_{:= I_y}.
\end{aligned}
$$
To evaluate $I_y$, we expand the quadratic terms and complete the square:
$$\begin{aligned}
	&\|y - y_j^{(a)}\|^2 + \|y - y_l^{(a)}\|^2\\ =& 2\|y\|^2 - 2y^\top(y_j^{(a)} + y_l^{(a)}) + \|y_j^{(a)}\|^2 + \|y_l^{(a)}\|^2 \\
	=& 2 \left( \|y\|^2 - y^\top(y_j^{(a)} + y_l^{(a)}) + \left\|\frac{y_j^{(a)} + y_l^{(a)}}{2}\right\|^2 \right) + \|y_j^{(a)}\|^2 + \|y_l^{(a)}\|^2 - 2\left\|\frac{y_j^{(a)} + y_l^{(a)}}{2}\right\|^2 \\
	=& 2 \left\| y - \frac{y_j^{(a)} + y_l^{(a)}}{2} \right\|^2 + \frac{1}{2}\|y_j^{(a)}\|^2 + \frac{1}{2}\|y_l^{(a)}\|^2 - (y_j^{(a)})^\top y_l^{(a)} \\
	=& 2 \left\| y - \frac{y_j^{(a)} + y_l^{(a)}}{2} \right\|^2 + \frac{1}{2}\|y_j^{(a)} - y_l^{(a)}\|^2.
\end{aligned}$$
Substituting this identity into $I_y$ gives
$$I_y = \exp\left(-\frac{\|y_j^{(a)} - y_l^{(a)}\|^2}{4\sigma_y^2}\right) \int \exp\left(-\frac{2\left\| y - \frac{y_j^{(a)} + y_l^{(a)}}{2} \right\|^2}{2\sigma_y^2}\right) \mathrm{d}y.$$
The remaining term is the integral of an unnormalized $d_y$ dimensional Gaussian kernel. Therefore,
$$\int \exp\left(-\frac{\left\| y - \frac{y_j^{(a)} + y_l^{(a)}}{2} \right\|^2}{\sigma_y^2}\right) \mathrm{d}y = (\sqrt{\pi}\sigma_y)^{d_y}.$$
Combining this result with the factor involving $x$ yields the closed form
$$\begin{aligned}
	\int \phi_j(x_i^{(a)}, y) \phi_l(x_i^{(a)}, y) \mathrm{d}y = \exp\left(-\frac{\|x_i^{(a)} - x_j^{(a)}\|^2 + \|x_i^{(a)} - x_l^{(a)}\|^2}{2\sigma_x^2}\right) (\sqrt{\pi}\sigma_y)^{d_y}\times \exp\left(-\frac{\|y_j^{(a)} - y_l^{(a)}\|^2}{4\sigma_y^2}\right).
\end{aligned}$$

%
%


\end{document}